\documentclass[journal]{IEEEtran}
\usepackage{xcolor,soul,framed} 
\usepackage{cite}
\usepackage{amsmath,amssymb,amsfonts}
\usepackage{graphicx}
\usepackage{textcomp}
\usepackage{dcolumn}
\usepackage{bm}
\usepackage{subfigure} 
\usepackage{pict2e}
\usepackage{epstopdf}
\usepackage{arydshln}
\usepackage{gensymb}
\usepackage{lipsum}

\newcommand{\Fig}[1]{Fig.\,{\ref{#1}}}
\newcommand{\Figs}[1]{Figs.\,{\ref{#1}}}

\newcommand{\Sec}[1]{Section\,\ref{#1}}

\newcommand{\jj}{\mathrm{j}}

\begin{document}

\onecolumn
\newpage
\thispagestyle{empty}

\textbf{Copyright:}
\copyright 2020 IEEE. Personal use of this material is permitted.  Permission from IEEE must be obtained for all other uses, in any current or future media, including reprinting/republishing this material for advertising or promotional purposes, creating new collective works, for resale or redistribution to servers or lists, or reuse of any copyrighted component of this work in other works.\\

\textbf{Disclaimer:} This work has been published in \textit{IEEE Transactions on Antennas and Propagation}. \\

Citation information: DOI 10.1109/TAP.2021.3070150

\newpage

\twocolumn

\setcounter{page}{1}

\title{Exploring the Potentials of the Multi-modal Equivalent Circuit Approach for Stacks of \mbox{2-D} Aperture Arrays}


\author{Antonio Alex-Amor,  Francisco Mesa, \textit{Fellow, IEEE}, Ángel Palomares-Caballero, Carlos Molero, \textit{Member, IEEE}, and Pablo Padilla 
\thanks{This work was supported by the Spanish Research and Development National Program under Projects TIN2016-75097-P, RTI2018-102002-A-I00, B-TIC-402-UGR18, TEC2017-84724-P, and the predoctoral grant FPU18/01965; by Junta de Andalucía under project P18-RT-4830.}
\thanks{A. Alex-Amor, A. Palomares-Caballero, C. Molero and P. Padilla  are with the Departamento de Teor\'{i}a de la Se\~{n}al, Telem\'{a}tica y Comunicaciones, Universidad de Granada, 18071 Granada, Spain (email: aalex@ugr.es, angelpc@ugr.es; pablopadilla@ugr.es.}
\thanks{A. Alex-Amor is also with the Information Processing and Telecommunications Center, Universidad Polit\'{e}cnica de Madrid, 28040 Madrid, Spain}
\thanks{F. Mesa is with the Microwaves Group, Department of Applied Physics 1,
Escuela Técnica Superior de Ingenieria Informatica, Universidad de Sevilla,
41012 Sevilla, Spain;  (e-mail: mesa@us.es)}
}

\markboth{} %
{Alex-Amor \MakeLowercase{\textit{et al.}}: Exploring the Potentials of the Multi-modal Equivalent Circuit Approach for Stacks of \mbox{2-D} Aperture Arrays}

\maketitle

\newcommand*{\bigs}[1]{\vcenter{\hbox{\scalebox{2}[8.2]{\ensuremath#1}}}}

\newcommand*{\bigstwo}[1]{\vcenter{\hbox{ \scalebox{1}[4.4]{\ensuremath#1}}}}

\begin{abstract}
Many  frequency  selective  surface (FSS) structures are based on the use of a single periodic array of slot/apertures in a conducting sheet embedded in a layered medium. However, it is well known that stacking several conducting sheets and breaking the alignment of the stack can bring multiple benefits to the structure.
In this paper, the analysis and design of stacks of \mbox{2-D} aperture arrays are carried out by exploiting  as much as possible all the potentialities of a rigorous and systematic formulation based on the multi-modal equivalent circuit approach (ECA). A key feature of the formulation is that linear transformations between the apertures of adjacent plates (rotation, translation, and scaling) can be dealt with from a purely analytical perspective.  This fact is of potential interest for many practical applications, such as the design of polarization converters, absorbers, filters, and thin matching layers. When the apertures have an arbitrary geometry, it can be applied a hybrid approach that combines the ability of commercial simulators to handle arbitrary geometries with the fast computation times and physical insight of the ECA. In general, either the purely analytical or the hybrid approach can be applied in those many practical scenarios where the spatial profile of the electric field on the considered apertures hardly changes with frequency. As an additional feature of the approach, the dispersion properties (phase/attenuation constants and Bloch impedance) of infinite periodic stacks can be derived and, in particular, analytical expressions for mirror- and glide-symmetric configurations are provided.

\end{abstract}


\IEEEpeerreviewmaketitle

\section{Introduction}

\IEEEPARstart{S}{tructured} surfaces have attracted a lot of attention both in microwaves~\cite{microwaves1, microwaves2, microwaves3, microwaves4}, THz~\cite{THz1, THz3} and the optical range~\cite{optical1, optical3} due to their versatility to control the reflection, refraction, and diffraction of the impinging waves by simply adjusting the geometrical parameters of the structure. This key feature has found multiple applications in science and engineering, such as FSS ~\cite{FSS1, FSS2, FSS3}, polarizers~\cite{polarizer1, polarizer2, Page-AP2018}, absorbers~\cite{THz1, absorber2, absorber3}, high-impedance surfaces~\cite{highimpedance1, highimpedance2} and electromagnetic bandgap (EBG) devices~\cite{EBG1, EBG3}.

A subclass of structured surfaces that are of particular interest in electromagnetism are one-dimensional~(1-D) strip/slit gratings~\cite{strip1, strip2} as well as two-dimensional~(\mbox{2-D}) periodic arrangements of metal patches and/or perforated apertures~\cite{apertures1, patches1, patches2, patches3, Elefth-AP2018,  Page-AP2020, Costa-AP2020, apertures3} in a layered medium. For these periodic structures, the scattering properties associated with an incident plane wave can be derived from a general waveguide discontinuity problem where periodic boundary conditions are applied~\cite{Collin, PBCproblem}. From this fact, it directly follows that the scattering problem can be analyzed in a rigorous manner from a circuit model perspective~\cite{Kurokawa, eca_magazine, ec1, ec8, ec10}. 

Most of the structures analyzed in the previous references consist of a single metal layer embedded in a layered dielectric medium. Nonetheless, it is well known that stacking several metal layers opens new possibilities to the design, such as the existence of transmission and rejection bands, increase of the operating bandwidth, appearance of negative-index refraction bands, enhanced performance of polarization converters, etc~\cite{Elefth-AP2018, stacked_benefits1, stacked_benefits2, stacked_benefits3}. Some equivalent circuits have been proposed to model the performance of stacked structures~\cite{heuristic1, heuristic2, Xu-MTT2018, Page-AP2018, Page-AP2020, Costa-AP2020}. However, some of these works (for instance, \cite{Xu-MTT2018, Page-AP2018, Page-AP2020, Costa-AP2020}) require a substantial assistance of previous full-wave simulations and the scope of some others (for instance, \cite{heuristic1, heuristic2})  is focused to very particular configurations and their formulation mainly based on a heuristic rationale. In these latter works, the proposed circuit models fail to take into account the strong coupling between the stacked layers when these are closely spaced. In the present work, we are interested in an equivalent circuit approach (ECA) that manages to include the coupling effects while offering a good physical insight on the scattering problem, and all of this by means of an analytical procedure~\cite{eca_magazine}. This kind of insightful works starts from basic electromagnetic principles and even/odd excitation techniques~\cite{evenodd1, evenodd2} to study aligned and symmetric configurations, which nonetheless is a limiting factor of the functionalities of the stacked structure. 

Recently, the range of use of these more rigorous equivalent circuits was successfully extended in~\cite{arbitrary2D} to model aligned stacks of apertures. Interestingly, breaking the alignment of the stacked metal layers can lead to enhanced performances of the stacked structure, especially with the inclusion of some of the revisited higher symmetries~\cite{highersymmetries2}. Glide symmetry is a kind of higher symmetry particularly useful in planar and stacked structures which involves a mirroring and a displacement of half a period between adjacent layers. Their effects cannot be modeled with the formulation presented in~\cite{arbitrary2D}, as the layers were required to be aligned along the vertical direction. The implementation of glide symmetry makes it possible to suppress the lowest stopband of the first propagating modes~\cite{glide_stopband1}, reduce the frequency dispersion of the structure~\cite{glide_dispersion1, glide_dispersion2, glide_dispersion3}, increase the equivalent refractive index~\cite{glide_index1, glide_index2}, and produce wideband anisotropy~\cite{glide_anisotropy2}. 

The above beneficial properties are expected to be efficiently analyzed with the extension of the ECA proposed in this work. To reach this goal, the multimodal equivalent-circuit methodology reported in~\cite{eca_magazine} is now extended to accurately compute the scattering properties of asymmetrical and nonaligned stacks formed by slot-based \mbox{2-D} periodic arrays with arbitrary apertures under normal and oblique incidences. The proposed formulation is a non-trivial extension of \mbox{1-D} case reported in~\cite{molero_asymmetrical1D}, which aims to explore the limits of applicability of the equivalent-circuit modelling by including  a second spatial dimension and by discussing new possibilities of analysis and/or applications.  Furthermore, stacked slot-based structures possessing glide symmetry can be studied using the present circuit perspective. This is a very appreciated feature since glide-symmetric structures can rarely be described by means of circuit models due to the strong and non-trivial interaction between adjacent layers~\cite{glide_dispersion3, glide_circuitmodel}. However, the strong couplings related to closely-spaced layers can be fully taken into account with the present approach. Thus, the proposed ECA will reveal itself as a very efficient tool for the design of wideband radomes, polarization converters, filters, absorbers, and many other devices based on stacked metallo-dielectric layers, even in complex scenarios.   

The work is organized as follows. Section\,\ref{sec:Analys} presents the formulation applied to the computation of the scattering properties of asymmetrical and nonaligned stacks of \mbox{2-D} arbitrarily shaped apertures. Section\,\ref{sec:Apertures} particularizes the study to canonical geometries, such as rectangular and annular apertures. Section\,\ref{sec:Arbitrary} analyzes apertures of arbitrary shape. Section\,\ref{sec:3-D} illustrates how to carry out a  dispersion analysis of 3-D periodic stacks for a frequency range only achieved before by commercial simulators. Finally, Section\,\ref{sec:Conc} summarizes the main conclusions extracted from the work.

\section{\label{sec:Analys} Analysis}

\subsection{\label{sec:deriv} Formal Derivation}

\begin{figure}[t]
	\centering
	\subfigure[]{\includegraphics[width=0.58\columnwidth]{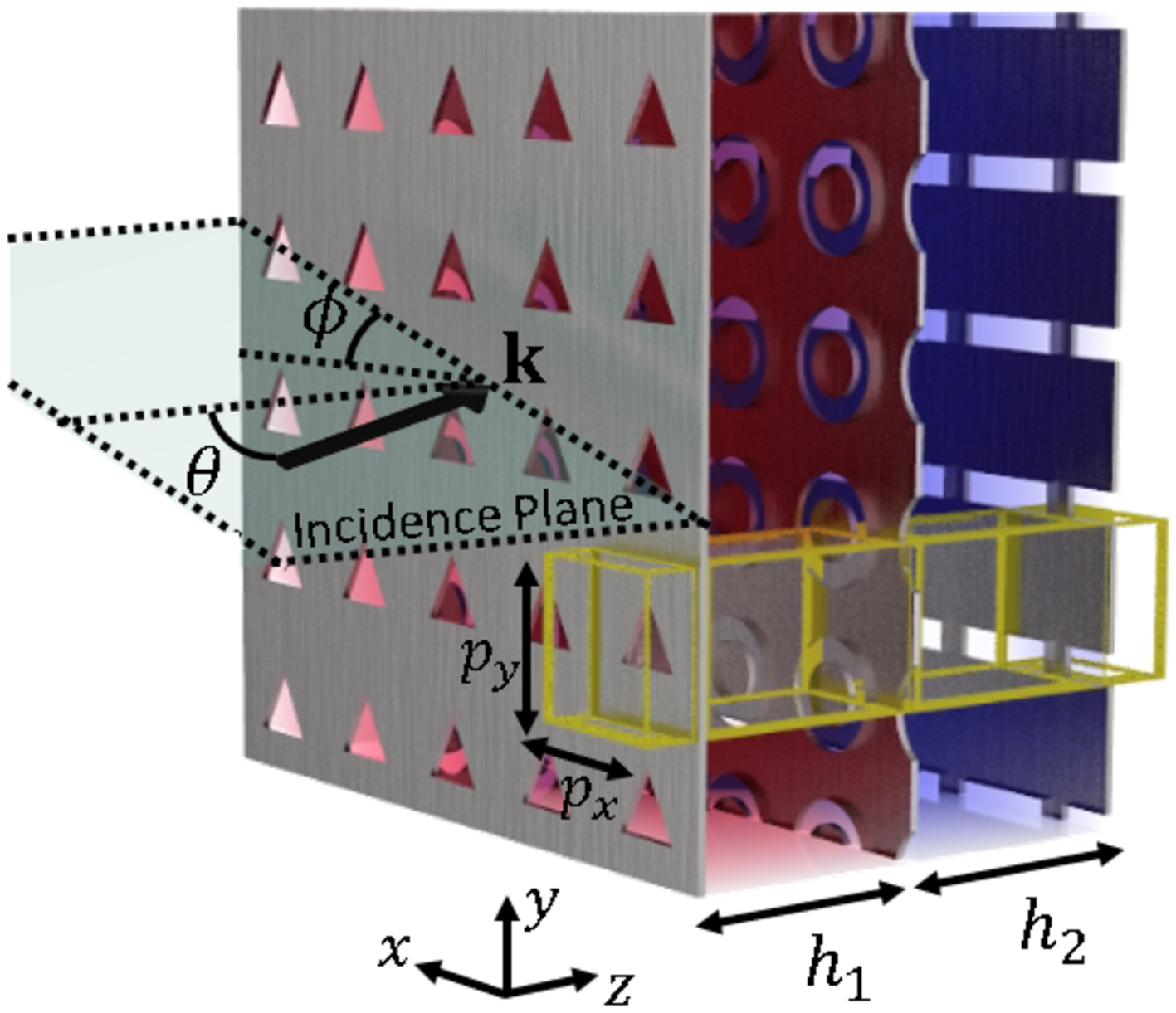}	} 
	\hspace*{-0.3cm}
	\subfigure[]{\includegraphics[width=0.40\columnwidth]{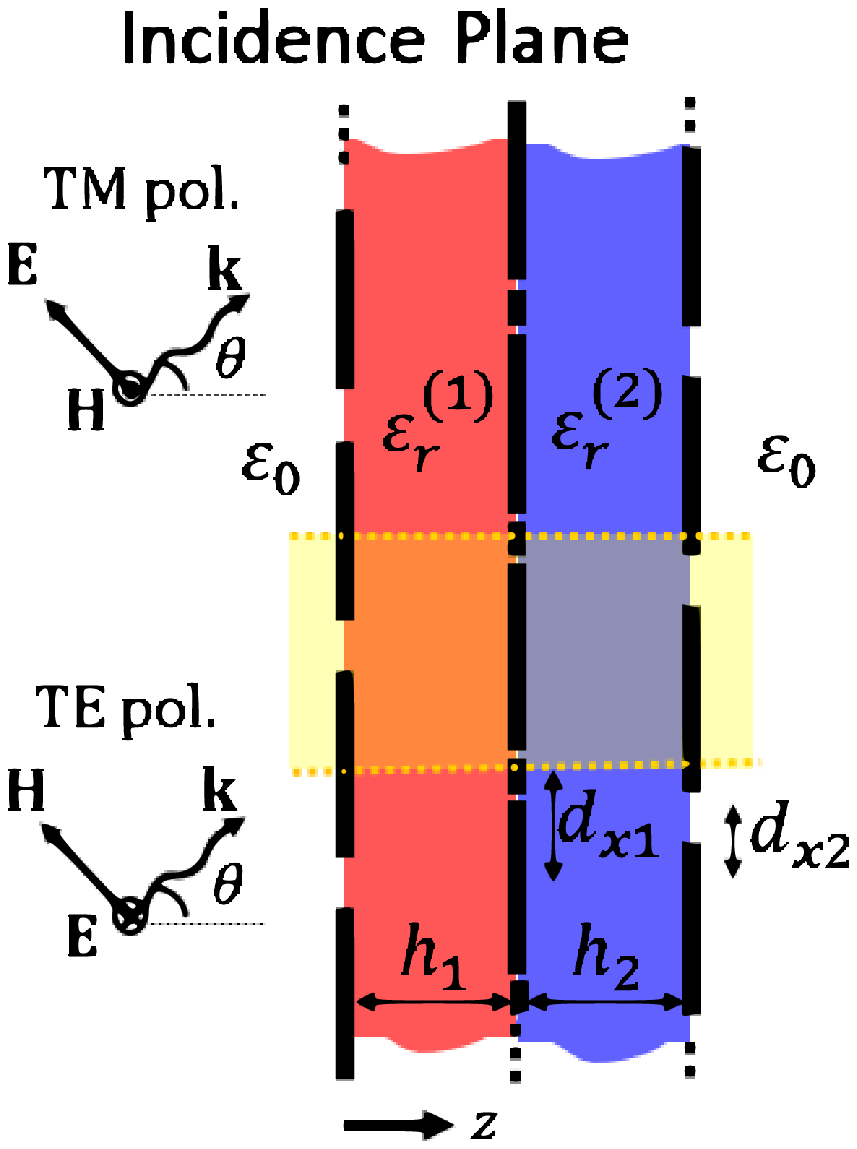}	} \\
	\subfigure[]{\includegraphics[width=\columnwidth]{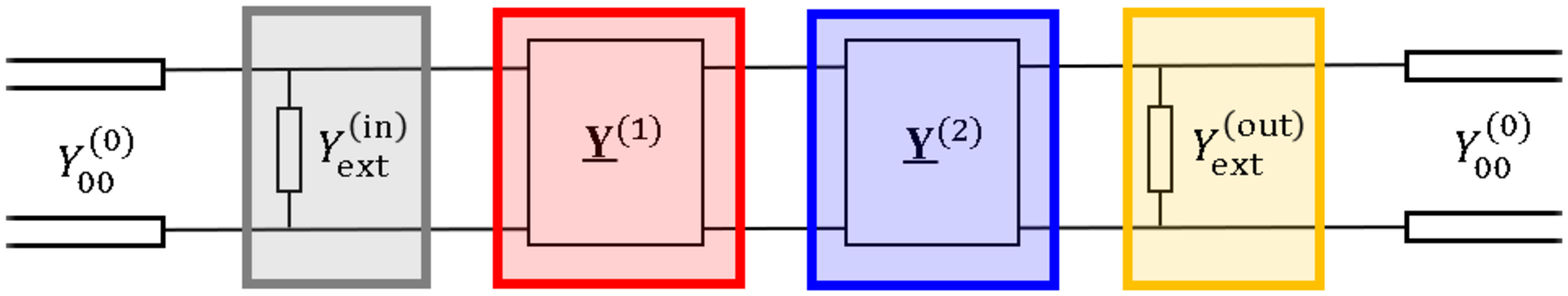}}
	\caption{(a)~Example of a stack of three asymmetrical and nonaligned \mbox{2-D} periodic arrays composed of arbitrary apertures and (b)~longitudinal view illustrating the incidence plane. (c)~General circuit model for the stack of three~arbitrary coupled apertures.} 
	\label{fig:struct-3lay}
\end{figure}
This section will first briefly outline the general procedure already reported in previous works of some of the authors to deal with a stack of $N$ metallic screen periodically perforated with arbitrary apertures. For simplicity, let us consider the stack of three nonaligned \mbox{2-D} periodic arrays composed of strongly-coupled arbitrary apertures displayed in~\Fig{fig:struct-3lay}(a), upon which a time-harmonic incident plane wave of angular frequency $\omega = 2 \pi f$ is obliquely impinging with a wavevector $\mathbf{k}_\text{inc}=(k_{x0},k_{y0},k_{z0})$ given by
\begin{align}
    k_{x0} & = \sqrt{\varepsilon_r^{(0)}} k_0 \sin\theta \cos\phi \\
    k_{y0} &= \sqrt{\varepsilon_r^{(0)}} k_0 \sin\theta \sin\phi \\
    k_{z0} &= \sqrt{\varepsilon_r^{(0)}} k_0 \cos\theta  
\end{align}
where $\varepsilon_r^{(0)}$ is the relative permittivity of the incident medium, $k_0$ is the vacuum wavenumber, and $\theta$ and $\phi$ are the elevation and azimuth angles of the incident wave, respectively. The adjacent metallic screens of the stack are separated with dielectrics of relative permittivity~$\varepsilon_{r}^\mathrm{(i)}$ and thickness~$h_i$. 

\begin{figure*}[t]
	\centering
\includegraphics[width=0.85\textwidth]{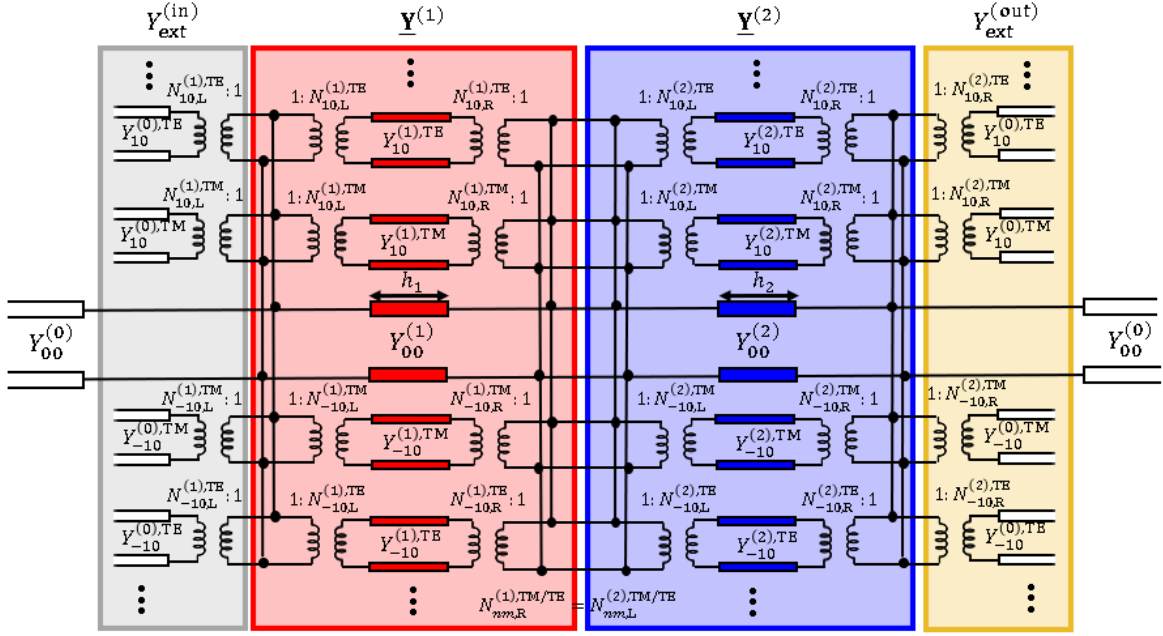}
	\caption{Detailed circuit model for a 3-layer asymmetrical and nonaligned stack of arbitrary apertures. Harmonics of different order are coupled together through the associated parallel-connected transmission lines loaded with transformers. } 
	\label{fig:coupled}
\end{figure*}

The tangential electric field on the aperture of a unit cell of one of the metal screens, $\mathbf{E}_t(x,y,\omega)$, is modeled in the following way:
\begin{equation}\label{eq:Et}
    \mathbf{E}_t(x,y,\omega) = F(\omega)\, \mathbf{E}_a(x,y) 
\end{equation}
where $F(\omega)$ is a frequency-dependent complex factor and $\mathbf{E}_a(x,y)$ is the assumed frequency-independent spatial profile. This assumption, key for the development of the formulation, is found to be applicable in a wide frequency band for many practical structures, even beyond the grating-lobe regime~\cite{eca_magazine, arbitrary2D}. The spatial profile $\mathbf{E}_a(x,y)$ can be expressed in closed form for canonical geometries (such as the rectangular and annular apertures considered in Appendix~A), which helps to reduce considerably the computational effort as it will be exploited in~\Sec{sec:Apertures}. For more complex aperture geometries, the use of any full-wave commercial software allows us to extract the spatial profile from the simulation of a single (non-stacked) free-standing metallic screen at just one particular frequency value. With this limited use of the full-wave simulator, we can combine the advantages of the ECA with the versatility of commercial software to deal with arbitrary geometries~\cite{patches2}. This advantageous assistance of commercial simulators will be exploited in~\Sec{sec:Arbitrary} for the analysis of nonaligned stacks of arbitrary geometry.

Based on the transfer (ABCD) matrix formalism~\cite{Collin,pozar}, the formulation derived in~\cite{molero_asymmetrical1D} for  nonaligned stacks of \mbox{1-D} periodic slit arrays is now extended to cover cases of \mbox{2-D}~apertures of arbitrary geometry. As clearly shown in~\cite[Fig.\,3(b)]{molero_asymmetrical1D}, the
circuit topology found for the stacks of slit/aperture arrays consists of blocks of parallel-connected transmission lines loaded with transformers. Thus,  as schematically shown in~\Fig{fig:struct-3lay}(b), the three-screen stacked structure in~\Fig{fig:struct-3lay}(a)  can be divided in four blocks: two associated with the so-called \textit{external} input and output regions, and other two regions associated with the \textit{internal} part of each pair of consecutive coupled arrays. The resulting network composed of four parallel-connected blocks is  depicted in~\Fig{fig:coupled}. At the light of this network, taking into account the parallel nature of the connections, the external input/output regions are completely characterized by the following single admittance:
\begin{multline}\label{eq:Yext}
    Y_\text{ext}^\text{(in)/(out)} = \sum_{\substack{n,m=-\infty \\ n,m\neq (0,0)}}^\infty \!\!  \Big[ \left(N_{nm,\text{L/R}}^{(1)/(2),\text{TM}}\right)^2  \,Y_{nm}^{(0),\text{TM}} \\+ 
    \left(N_{nm,\text{L/R}}^{(1)/(2),\text{TE}} \right)^2 \,Y_{nm}^{(0),\text{TE}} \Big]
\end{multline}
where the index $(1)/(2)$ refers to the $(i)$-th internal region and the index~$(0)$ refers to free space. The network topology of the internal $(i)$-th block  is formed by the parallel-connected transmission lines associated with the harmonics corresponding with the dielectric layers inside the corresponding pair of coupled arrays  and the transformers at the left and right hand sides of these transmission lines. The internal regions can then be modelled by the following admittance matrix ($i=1,2$):
\begin{equation}
\underline{\mathbf{Y}}^{(i)}=
    \begin{bmatrix}
        Y_{11}^{(i)} & Y_{12}^{(i)} \\
        Y_{21}^{(i)} & Y_{22}^{(i)} 
\end{bmatrix}
\end{equation}
the entries of which are calculated as ($u,v=1,2$)
\begin{equation} \label{eq:Yuv} 
    Y_{uv}^{(i)}=\sum_{n,m=-\infty}^\infty \!\! (Y_{uv,nm}^{(i),\mathrm{TM}} + Y_{uv,nm}^{(i),\mathrm{TE}})  
\end{equation}
with (TX will stand indistinctly for either TM or TE)
\begin{align} 
\label{Y11nm}
    Y_{11,nm}^{(i),\text{TX}} &= \left({N_{nm,\text{L}}^{(i),\text{TX}}}\right)^2 \left[ -\jj Y_{nm}^{(i),\text{TX}}\cot(k_{z,nm}^{(i)}h_i)\right] 
\\ \label{Y12nm}
 Y_{12,nm}^{(i),\text{TX}} & = N_{nm,\text{L}}^{(i),\text{TX}}
\,N_{nm,\text{R}}^{(i),\text{TX}} \left[ \jj Y_{nm}^{(i),\text{TX}} \csc(k_{z,nm}^{(i)}h_i)\right]
\end{align}
\begin{align} 
     Y_{21,nm}^{(i),\text{TX}} &=   N_{nm,\text{R}}^{(i),\text{TX}}\,{N_{nm,\text{L}}^{(i),\text{TX}}} 
      \left[\jj Y_{nm}^{(i),\text{TX}} \csc(k_{z,nm}^{(i)}h_i)\right]
\\ \label{Y22nm}
     Y_{22,nm}^{(i),\text{TX}} & = 
    \left(N_{nm,\text{R}}^{(i),\text{TX}}\right)^2 \left[ -\jj Y_{nm}^{(i),\text{TX}} \cot(k_{z,nm}^{(i)}h_i)\right]\,.
\end{align}
In the above derivations, it has been assumed that the internal region~$(i)$ only comprises a single dielectric~$(i)$. As reported in~\cite{patches1}, if the internal region~$(i)$ is composed of several dielectric layers, we should substitute the transmission lines associated with the harmonics inside the single dielectric by the corresponding cascade of transmission lines that accounts for the layered environment (namely, the terms inside the brackets in the above expressions should be substituted by the corresponding ones associated with the cascade of dielectric layers). 
The indexes~L and~R in \eqref{eq:Yext} and \eqref{Y11nm}--\eqref{Y22nm} refer respectively to the left-side and right-side aperture arrays that bound the dielectric~($i$). $Y_{nm}^{(i),\text{TX}}$ is the wave admittance of the $(n,m)$-th (TX $\equiv$ TM/TE) harmonic at dielectric~$(i)$, defined as
\begin{gather}
    \label{eq:YTM}
    Y_{nm}^{(i),\text{TM}}= \dfrac{1}{\eta^{(i)}}
                                   \dfrac{ k^{(i)} }{ k_{z,nm}^{(i)} }  \\
    \label{eq:YTE}
    Y_{nm}^{(i),\text{TE}}= \dfrac{1}{\eta^{(i)}}
                                   \dfrac{ k_{z,nm}^{(i)} }{ k^{(i)} }  
\end{gather}
with $\eta^{(i)}$ being the wave impedance of the $i$-th medium and $k_{z,nm}^{(i)}$ the longitudinal wavenumber of the $(m,n)$-th harmonic in such medium, given by
\begin{equation}\label{kzmn}
    k_{z,nm}^{(i)}=\sqrt{[k^{(i)}]^2-|\mathbf{k}_{t,nm}|^2}\;.
\end{equation}
In \eqref{kzmn}, $k^{(i)}=\sqrt{\varepsilon_r^{(i)}} k_0$ and  $\mathbf{k}_{t,nm}$ is its associated transversal wavevector, expressed as 
\begin{equation}
    \mathbf{k}_{t,nm}=k_{xn}\mathbf{\hat{x}} + k_{ym}\mathbf{\hat{y}} = (k_{x0} + k_{n}) \mathbf{\hat{x}} + (k_{y0} + k_{m}) \mathbf{\hat{y}}
\end{equation}
with
\begin{equation*}
    k_n=\frac{2\pi n}{p_x} \qquad ,\qquad 
    k_m=\frac{2\pi m}{p_y}
\end{equation*}
and $p_x$, $p_y$ the periods of the unit cell in the~$x$ and $y$~directions.


From a circuit standpoint, the coefficients $N_{nm,\text{L}/\text{R}}^{(i),\text{TX}}$ are the turn ratios of transformers associated with the \mbox{$(m,n)$-th} harmonics in region~(i)~\cite{patches1}. Mathematically they stand for the projection of the $(n,m)$-th harmonic on the 2-D Fourier transform of the spatial profile $\mathbf{E}^{(i)}_{a,\text{L/R}}(x,y)$  at the corresponding left-/right-side apertures; that is,
\begin{align}
    \label{eq:NnmTM} 
    N_{nm,\text{L/R}}^{(i),\text{TM}} & =  \widetilde{\mathbf{E}}_{a,\text{L/R}}^{(i)}(\mathbf{k}_{t,nm}) \cdot \hat{\mathbf{k}}_{t,nm} \\
    \label{eq:NnmTE} 
    N_{nm,\text{L/R}}^{(i),\text{TE}} & =  \widetilde{\mathbf{E}}_{a,\text{L/R}}^{(i)}(\mathbf{k}_{t,nm}) \cdot 
   (\hat{\mathbf{k}}_{t,nm} \times \mathbf{\hat{z}})
\end{align}
where $\hat{\mathbf{k}}_{t,nm}$ is the unit vector associated with $\mathbf{k}_{t,nm}$ and $\widetilde{\mathbf{E}}_{a,\text{L/R}}^{(i)}(\mathbf{k}_{t,nm})$ is the \mbox{2-D} Fourier transform of the spatial profile in the left/right apertures of the coupled arrays~$(i)$ calculated at~$\mathbf{k}_{t,nm}$.

In general, $N_{nm,\text{L}}^{(i),\text{TX}}$ and $N_{nm,\text{R}}^{(i),\text{TX}}$ have different and unrelated values, since $\widetilde{\mathbf{E}}_{a,\text{L}}^{(i)}(\mathbf{k}_{t,nm})$ can be arbitrarily different from $\widetilde{\mathbf{E}}_{a,\text{R}}^{(i)}(\mathbf{k}_{t,nm})$. However, in many practical situations, the left and right apertures can be related by simple algebraic transformation (translation, rotation, reflection and dilation), which then makes it possible to also find simple algebraic relations between the involved Fourier transforms. As an example, the misalignment of consecutive periodic arrays is taken into account by means of
\begin{equation}\label{eq:EREL}
      \widetilde{\mathbf{E}}_{a,\text{R}}^{(i)}(\mathbf{k}_{t,nm}) = 
      \widetilde{\mathbf{E}}_{a,\text{L}}^{(i)}(\mathbf{k}_{t,nm})
      \mathrm{e}^{\,\jj \mathbf{k}_{t,nm}\cdot \mathbf{d}}
\end{equation}
where $\mathbf{d}=d_x\mathbf{\hat{x}} + d_y\mathbf{\hat{y}}$ represents the displacement of the aperture (this misalignment was already considered in~\cite{molero_asymmetrical1D} for \mbox{1-D} periodic arrays). The case of stacks of rotated periodic arrays will be treated in~\Sec{sec:Nonaligned}. An interesting case raises when $\mathbf{d}=p_x/2\mathbf{\hat{x}} + p_y/2\mathbf{\hat{y}}$; namely, when the periodic stack has glide symmetry~\cite{highersymmetries2,glide_circuitmodel}. Under normal incidence, it means that \eqref{eq:EREL} turns into 
\begin{equation}\label{eq:EREL_glide}
      \widetilde{\mathbf{E}}_{a,\text{R}}^{(i)}(\mathbf{k}_{t,nm}) = 
      (-1)^{n+m} \,
      \widetilde{\mathbf{E}}_{a,\text{L}}^{(i)}(\mathbf{k}_{t,nm})
\end{equation}
which implies that
\begin{equation}\label{eq:NRNL_glide}
  N_{nm,\text{R}}^{(i),\text{TM}} =  (-1)^{n+m} \, N_{nm,\text{L}}^{(i),\text{TM}}\;.
\end{equation}
This particular relation between the transformer ratios is in very close correspondence with the discussion in~\cite{GuidoMM} on the symmetry of the even/odd harmonics when dealing with glide-symmetric structures. Thus, it is found that each $(n,m)$~harmonic involves the presence of a magnetic/electric wall in the middle plane of the sub-unit cell depending on whether~$n+m$ is even/odd. This interesting feature of Bloch modes in glide-symmetric structures is key for providing many of the beneficial properties of these periodic structures~\cite{JMW2021}. 

It should be noted that all the previous expressions from \eqref{eq:Yuv} to~\eqref{eq:NnmTE} are frequency dependent. Thus, the double sum in~\eqref{eq:Yuv} would have to be performed for every frequency value in an eventual frequency sweeping. However, for high-order (ho) harmonics ($k_{xn}^2 + k_{ym}^2 \gg \varepsilon_r k_0^2$), it is apparent that the wavenumber and wave admittances can be well approximated as~\cite{ec8, patches1}
\begin{equation}
      k_{z,nm}^{(i),\mathrm{ho}} \approx -\jj\alpha_{nm}= -\jj \sqrt{k_n^2 + k_m^2}
\end{equation}
and
\begin{equation}
     Y_{nm}^{(i),\mathrm{ho}} \approx 
     \begin{cases}
         \dfrac{\jj\omega \varepsilon_0 \varepsilon_r^{(i)}}{\alpha_{nm}} \equiv \jj\omega C_{nm}^{(i)}\,, & \text{TM harmonics} \\[2ex] 
         \dfrac{\alpha_{nm}}{\jj\omega\mu_0} \equiv \dfrac{1}{\jj\omega L_{nm}}\,, & \text{TE harmonics}\,. \\
    \end{cases}
\end{equation}
It implies that a great deal of computational effort can be saved in the computation of ~\eqref{eq:Yuv} by splitting the double infinite sum  into a low order (lo) contribution ($|n,m|\leq N$), which is frequency dependent but only comprises a few terms, plus a higher order (ho) contribution ($|n,m|\geq N+1$) that is frequency independent; namely,
\begin{multline} \label{Yuvho}
        Y_{uv}^{(i)}(\omega)=\sum_{n,m=-N}^N \left[Y_{uv,nm}^{(i),\mathrm{TM,lo}}(\omega) + Y_{uv,nm}^{(i),\mathrm{TE,lo}}(\omega) \right]\\
    +    \sum_{|n,m|\geq N+1}^\infty \left[Y_{uv,nm}^{(i),\mathrm{TM,ho}} + Y_{uv,nm}^{(i),\mathrm{TE,ho}} \right]\;.
\end{multline}
Therefore, the computational effort in the frequency-sweeping computation of each of the admittance matrices $\underline{\mathbf{Y}}^{(i)}(\omega)$  lies almost entirely in the obtaining of the reduced summation associated with  $\underline{\mathbf{Y}}^\mathrm{lo}(\omega)$, with $\underline{\mathbf{Y}}^\mathrm{ho}$ needed to be computed just once and stored for subsequent use. In most of the cases studied in this paper, it suffices to take $N\lesssim 6$, although this value can be smaller if (i)~the upper frequency of analysis is not close to the onset of the diffraction limit ($f<c/p$, with $c$ being the speed of light), and (ii)~when the periods of the unit cell are electrically small; feature that is found in many applications of metasurfaces.

For a straightforward computation of the scattering parameters, the admittances $Y_\text{ext}^\text{(in)/(out)}$ associated with the external regions and the $\underline{\mathbf{Y}}^{(i)}$ admittance matrices associated with the internal regions are converted to transfer (ABCD) matrices~\cite{pozar}. Given the particular topology found for the equivalent circuit of a generic stack of $M\geq 2$ coupled layers, the problem can be split up into $M-1$ internal blocks plus 2~additional external blocks. Thus, the complete network shown in~\Fig{fig:coupled} can be represented in terms of a single transfer matrix computed as the product of the resulting four individual transfer matrices. It should pointed out that the final transfer matrix does contain all the relevant information about the propagating mode and all higher-order harmonics, as well as the possible couplings between them.

\subsection{\label{sec:cons} Additional Considerations}

Periodic arrays of patch-like scatterers can be regarded as complementary 
to aperture-like ones and, thus, the application of a similar procedure as 
above to patch-like arrays would lead to a complementary equivalent circuit topology.
In~\cite{patches1} it was shown that the equivalent-circuit topology of a single periodic 
array of patches has all the individual transmission lines associated 
with the different harmonics  connected in \textit{series}, unlike the \textit{parallel} 
configuration found for aperture-like arrays. If two or more 
patch-like arrays are to be stacked, the corresponding transmission lines 
associated with harmonics of the same order would have to be connected, as done 
in~\Fig{fig:coupled} for the aperture problem. However, the resulting patch-like 
connection gives rise to a very complicated network that, to the authors' knowledge, 
cannot be simplified in a similar fashion as in the concatenation of simple 
$\underline{\mathbf{Y}}^{(i)}$ blocks shown in~\Fig{fig:coupled}. 
The lack of such straightforward connection complicates the mathematical treatment 
of the problem enormously and makes it very difficult to have a simple and insightful 
physical understanding of the structure from a circuit model standpoint. For this reason, 
stacks of patch-like arrays are out of the scope of this paper.

A relevant issue concerning the application of the ECA previously proposed
is the discussion of its limits of validity. As already  mentioned and previously reported 
in~\cite{limits}, the most relevant theoretical limitation of the approach comes from the 
validity of assumption~\eqref{eq:Et}; namely, that the spatial profile of the tangential field in the apertures does not vary too much with frequency. From a practical point of view, this assumption can be found satisfactory up to frequencies below the second ``excitable'' resonance of the aperture. Thus, for the case of an array of rectangular apertures of size $a\times b$ ($a>b$) with a normally incident electric field directed along the shorter dimension, the second excitable resonance will occur when $a\approx 3\lambda/2$; that is, for frequencies satisfying $f \lesssim 3c/(2a)$ (which includes a large frequency range well inside the diffraction regime).  In the case of oblique incidence, the limiting frequency can reduce to $f \lesssim c/a$ since the second excitable resonance may occur at $a\approx \lambda$. For other non-canonical geometries of the aperture such as the Jerusalem cross, the second excitable resonance may appear close to the first one, which would certainly reduce the range of applicability of the present approach. However, despite this fact, it can be stated that the ECA is found to work satisfactorily for many practical cases where the numerically-intensive full-wave approach can be advantageously substituted by the much simpler ECA.  Actually, this consideration is one of the main goals of the present work, where we explore different scenarios that might be thought to be intractable by means of the present quasi-analytic ECA.

\begin{figure}[t]
	\centering
	\subfigure[]{\includegraphics[width=0.4\columnwidth]{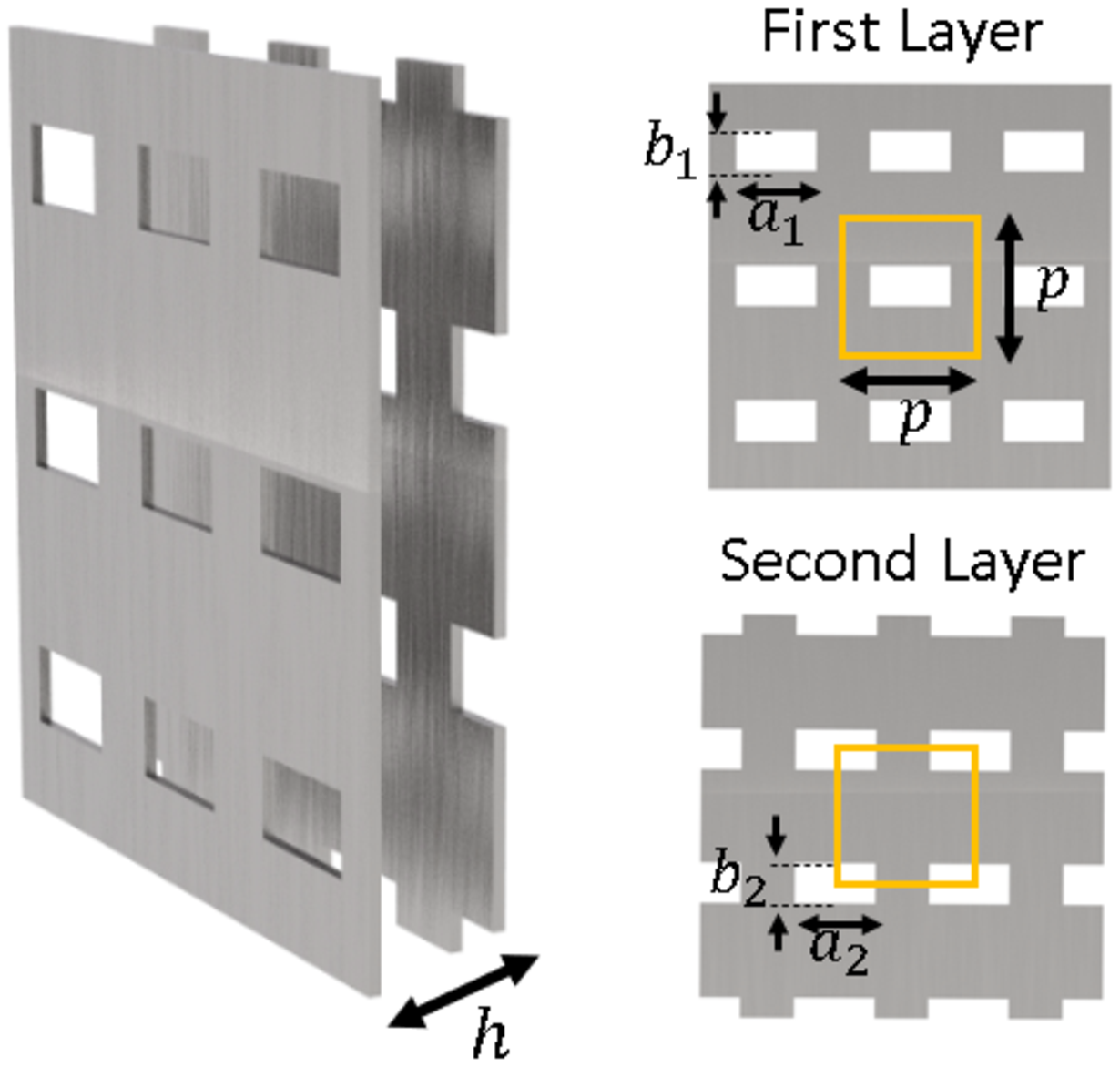}	}\,
	\subfigure[]{\includegraphics[width=0.49\columnwidth]{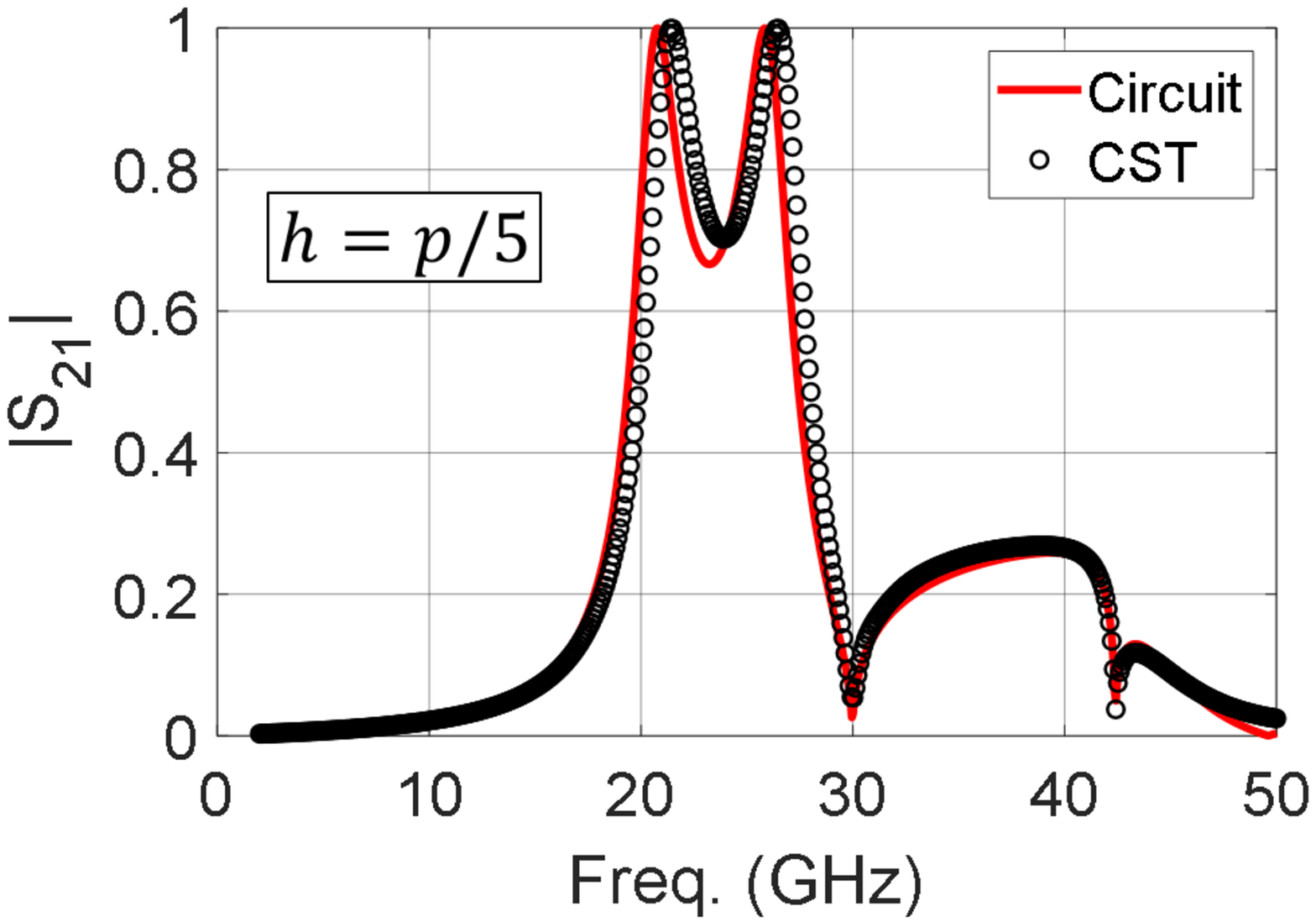}	}\\ 
	\hspace*{-0.3cm}
	\subfigure[]{\includegraphics[width=0.51\columnwidth]{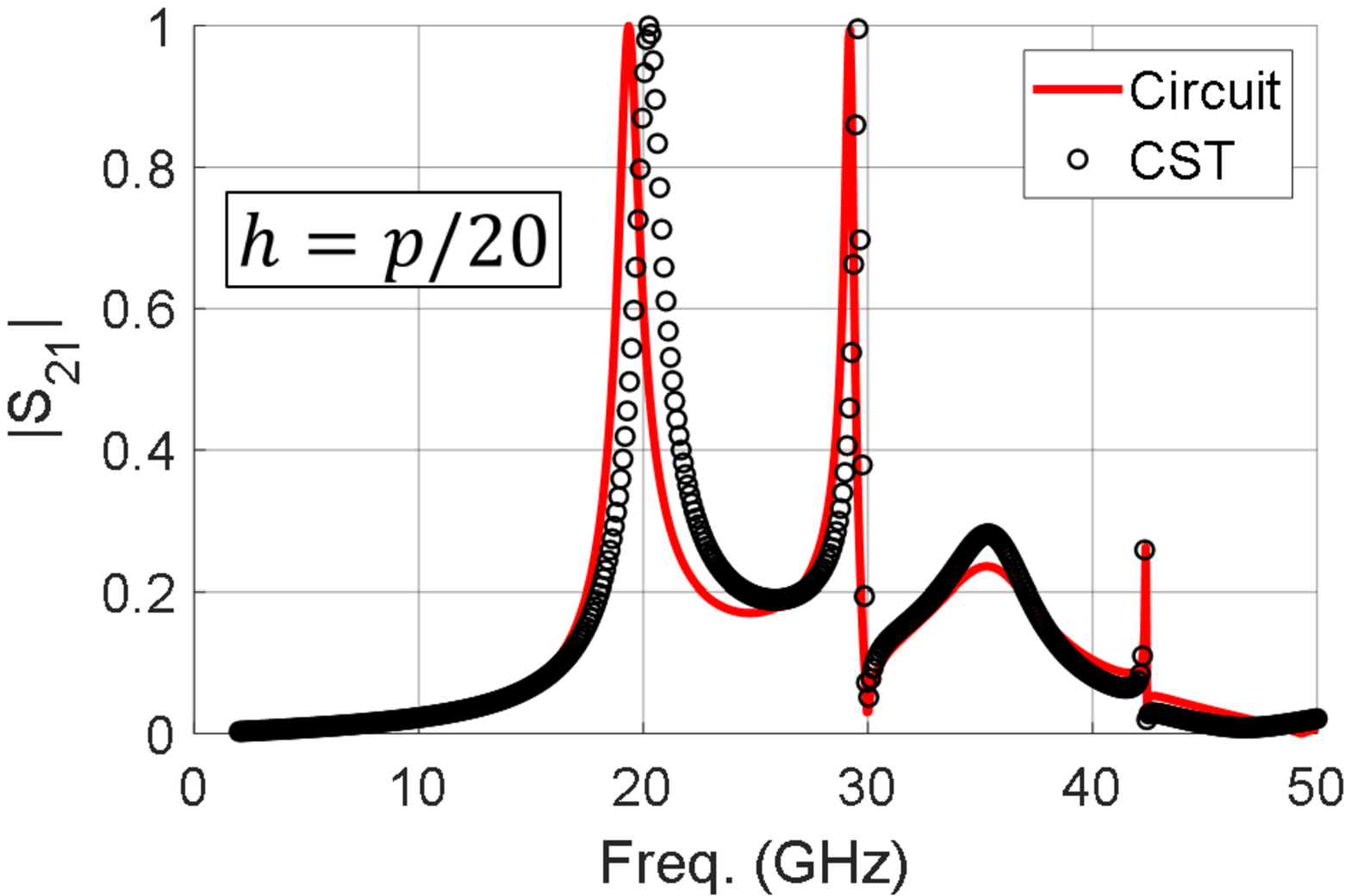}	} 
	\hspace*{-0.5cm}
	\subfigure[]{\includegraphics[width=0.51\columnwidth]{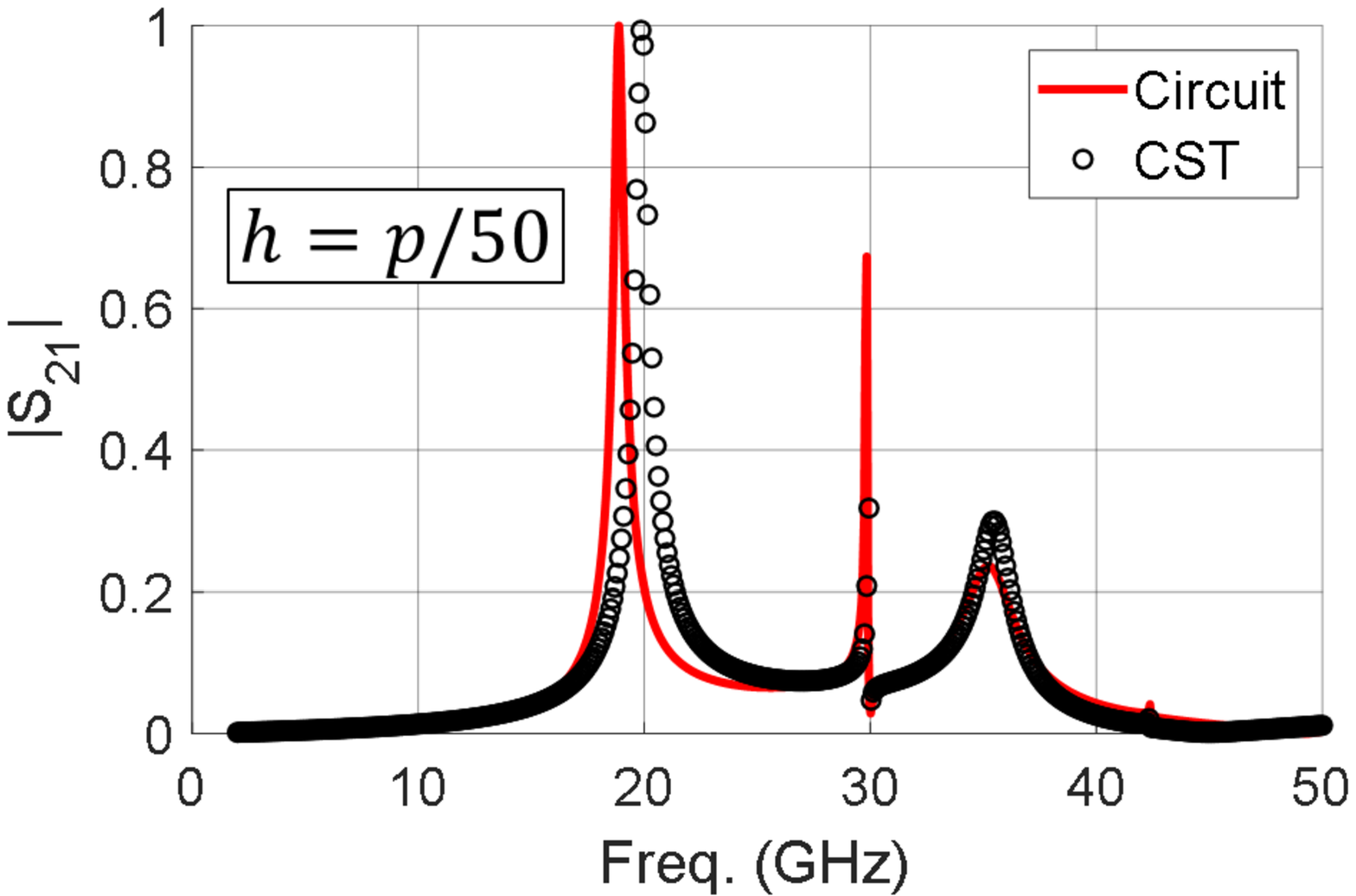}	} \\
	\caption{(a) Pair of strongly coupled arrays with rectangular apertures arranged in a glide-symmetric configuration. Transmission coefficient for a separation between arrays of (b)~$h=p/5$, (c)~$h=p/20$, and (d)~$h=p/50$. TM normal incidence is assumed. Geometrical parameters of the unit cell: $a_1=a_2=6$\,mm, $b_1=b_2=3$\,mm, $p_x=p_y=p=10$\,mm,  and $\varepsilon_r=1$. 
	} 
	\label{fig:closely_spaced}
\end{figure}

Another limitation discussed in~\cite{limits} concerns the variation of the spatial profile in the different arrays of the stack when the apertures are strongly coupled. That possible variation might be a relevant limiting factor in many practical cases since we are implicitly assuming that all the apertures in the stack have the same spatial profile as the one corresponding to each aperture taken isolated. In order to assess the relevance of this limitation, we will compare our ECA results with those provided by \textit{CST} in some cases of strong coupling for the configuration shown in \Fig{fig:closely_spaced}(a). In our simulations with commercial software \textit{CST}, we select the Frequency solver, configured with a maximum number of 20 cells per box model in the tetrahedral mesh, a maximum number of passes (finer mesh per iteration) of six, and 60 Floquet harmonics. Figs.\,\ref{fig:closely_spaced}(b)-(d)  show the transmission coefficient of a pair of strongly coupled arrays with rectangular apertures arranged in a glide-symmetric configuration when the separation between the arrays are $p/5$, $p/20$ and $p/50$, respectively (these cases correspond to $h=\lambda/5$, $h=\lambda/20$ and $h=\lambda/50$, taking $\lambda$ at the onset of the diffraction regime; namely, $f=30\,$GHz). The good agreement found between our ECA data with the ones provided by \textit{CST} in all these cases makes it apparent that our assumption of taking the ``isolated'' spatial profile for the apertures work reasonably well, even in the extreme scenario considered in  \Fig{fig:closely_spaced}(d).

\section{\label{sec:Apertures} Canonical Apertures}

In this section, the proposed ECA is used and tested to compute the scattering properties of stacked structures formed by apertures for which the spatial dependence of their tangential electric fields can be expressed by closed-form expressions. 

\subsection{Symmetrical and Aligned Stacks}

\begin{figure}[!h]
	\centering
	\subfigure[]{\includegraphics[width= 0.7\columnwidth]{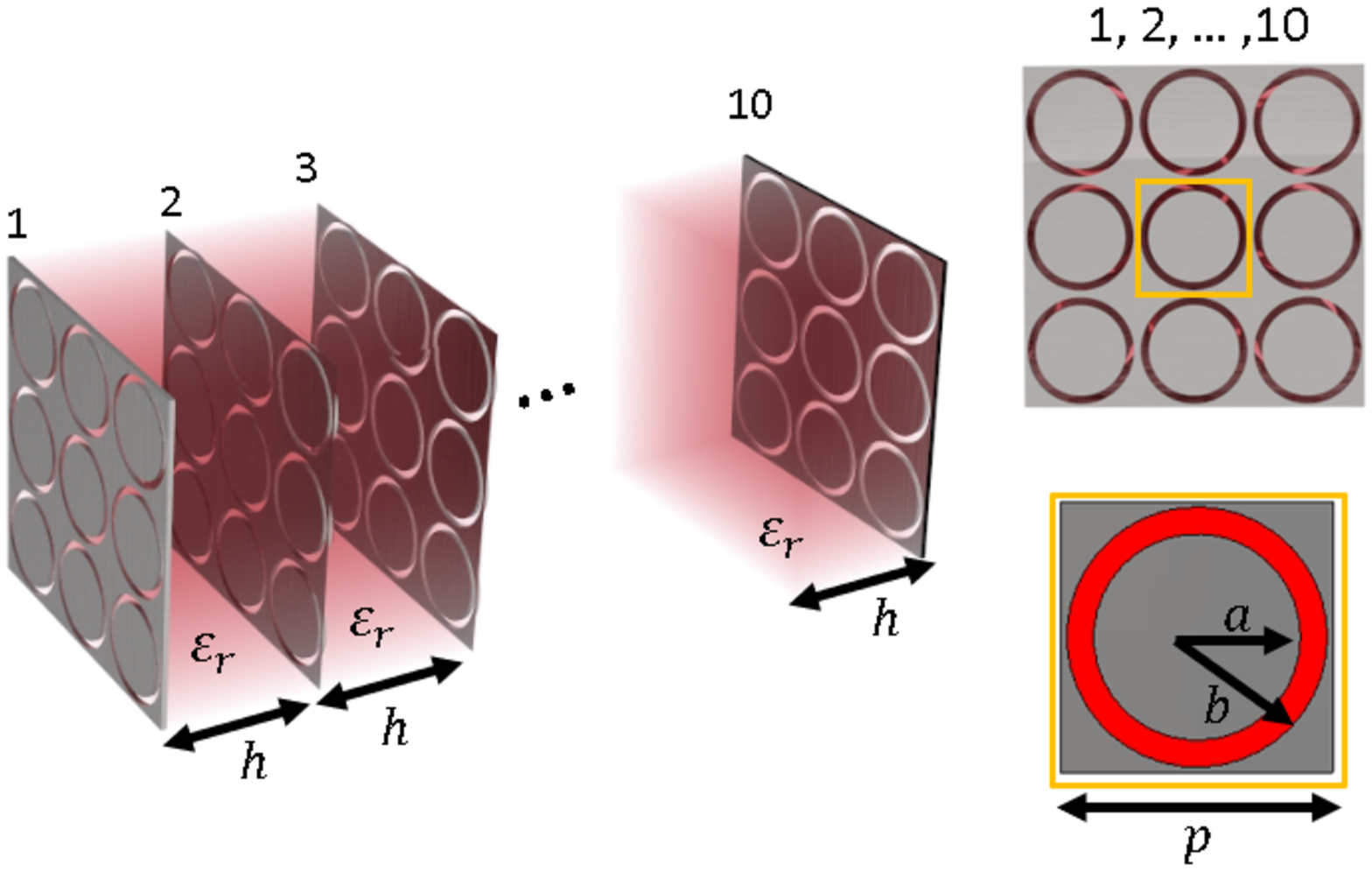}
	} 
	\subfigure[]{\includegraphics[width= 0.8\columnwidth]{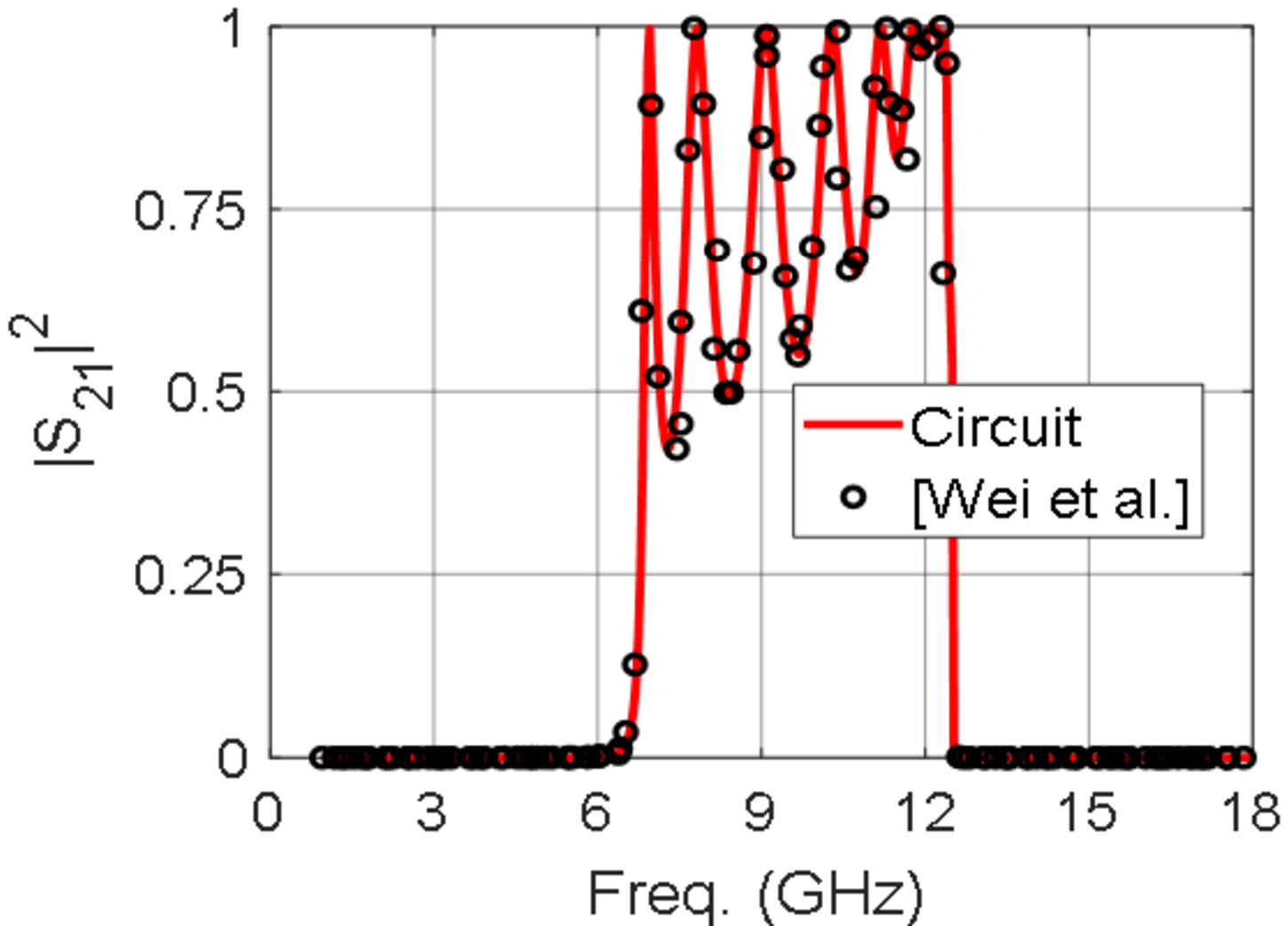}
	} 
	\caption{(a) Symmetrical and aligned stack of ten \mbox{2-D} arrays formed by annular apertures presented in~\cite{annular_stack}. (b)~Transmissivity versus frequency for TM~normal incidence. Geometrical parameters of the unit cell: $a=3.8$\,mm, $b=4.8$\,mm, $p_x=p_y=p=10$\,mm,  $h=1.575$\,mm, and $\varepsilon_r=2.65$.  }
	\label{fig:10lay}
\end{figure}

As a first study case, it is considered the multilayered structure studied in~\cite{annular_stack} and shown in~\Fig{fig:10lay}(a). The stack is made up by ten perfectly aligned metallic screens of periodic annular apertures separated by a dielectric of permittivity $\varepsilon_r$. The metal is assumed to be a perfect electric conductor (PEC) and no losses are considered in the dielectrics in this stage. High transmission can be achieved by stacking and alternating identical metallic and dielectric layers. This is appreciated in~\Fig{fig:10lay}(b), where the transmissivity is computed with the proposed approach and then compared with the original results in~\cite{annular_stack}.  For the computation, the electric field in the aperture is assumed to be well modeled by the function~\eqref{eq:Ea_anular} ($l=1$) in the Appendix, $N=5$ has been considered in~\eqref{Yuvho} and the double infinite sum has been truncated to $N_\mathrm{max}=10$. A good agreement is found between our closed-form results and the set of data in~\cite{annular_stack}. The good agreement is somewhat expected since the frequency range analyzed covers the range of validity of our approach discussed in Sec.\,\ref{sec:cons} (in this case, $f\lessapprox 15\,$GHz).  As the number of stacked metallic layers increases, more transmission peaks appear in the transmission spectra. For the structure under consideration, a passband emerges from 6.5\,GHz to 12.5\,GHz with transmissivity values over 50\%. It should be mentioned that all terms involved in the computation of our data are known in closed-form expressions, which allows us to carry out the study shown in the figure with a very reduced computational effort.

\subsection{\label{sec:Nonaligned} Glide-symmetric Stacks}

As previously discussed, breaking the alignment between two consecutive layers can bring some advantages to conventional FSSs. In particular, next we will study the effect of the introduction of glide symmetry in the structure previously analyzed. The half-period displacement implicit in the glide symmetry is taken into account here in a fully-analytical form. To the authors' knowledge, this is the first reported case where equivalent circuits can accurately model the strong higher-order coupling between Floquet harmonics in glide-symmetric FSS structures.

\begin{figure}[t]
	\centering
	\subfigure[\hspace*{-2.5cm}]{\includegraphics[width= 0.7\columnwidth]{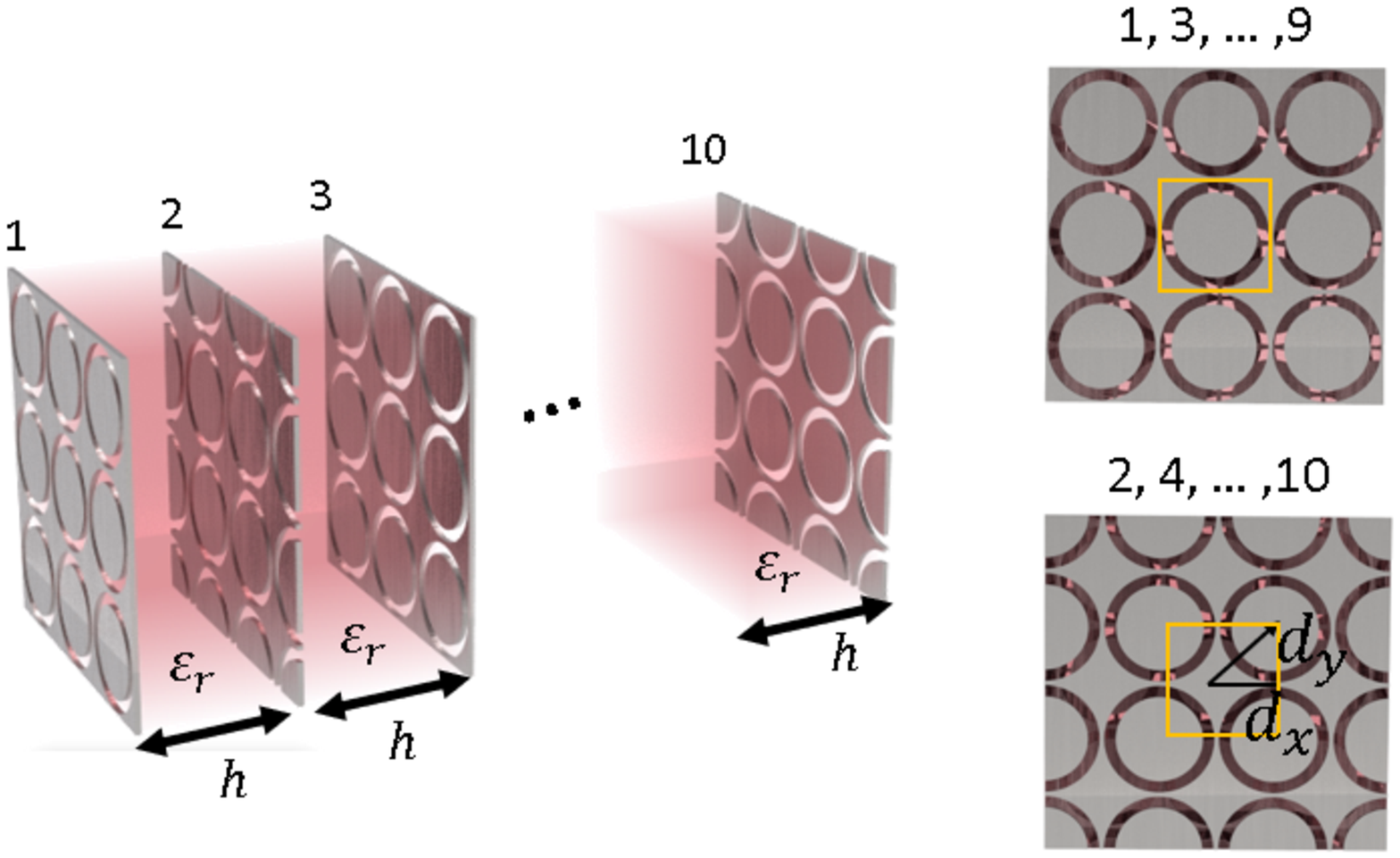}
	} 
	\hspace*{0.3cm}
	\subfigure{\raisebox{4ex}{\includegraphics[width= 0.17\columnwidth]{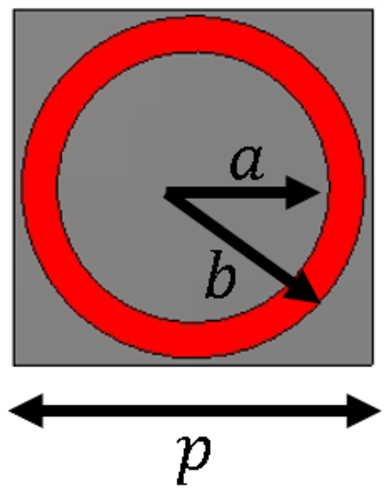}
	}} \\
	\setcounter{subfigure}{1}
	\subfigure[	\hspace*{-0.4cm}]{\includegraphics[width= 0.75\columnwidth]{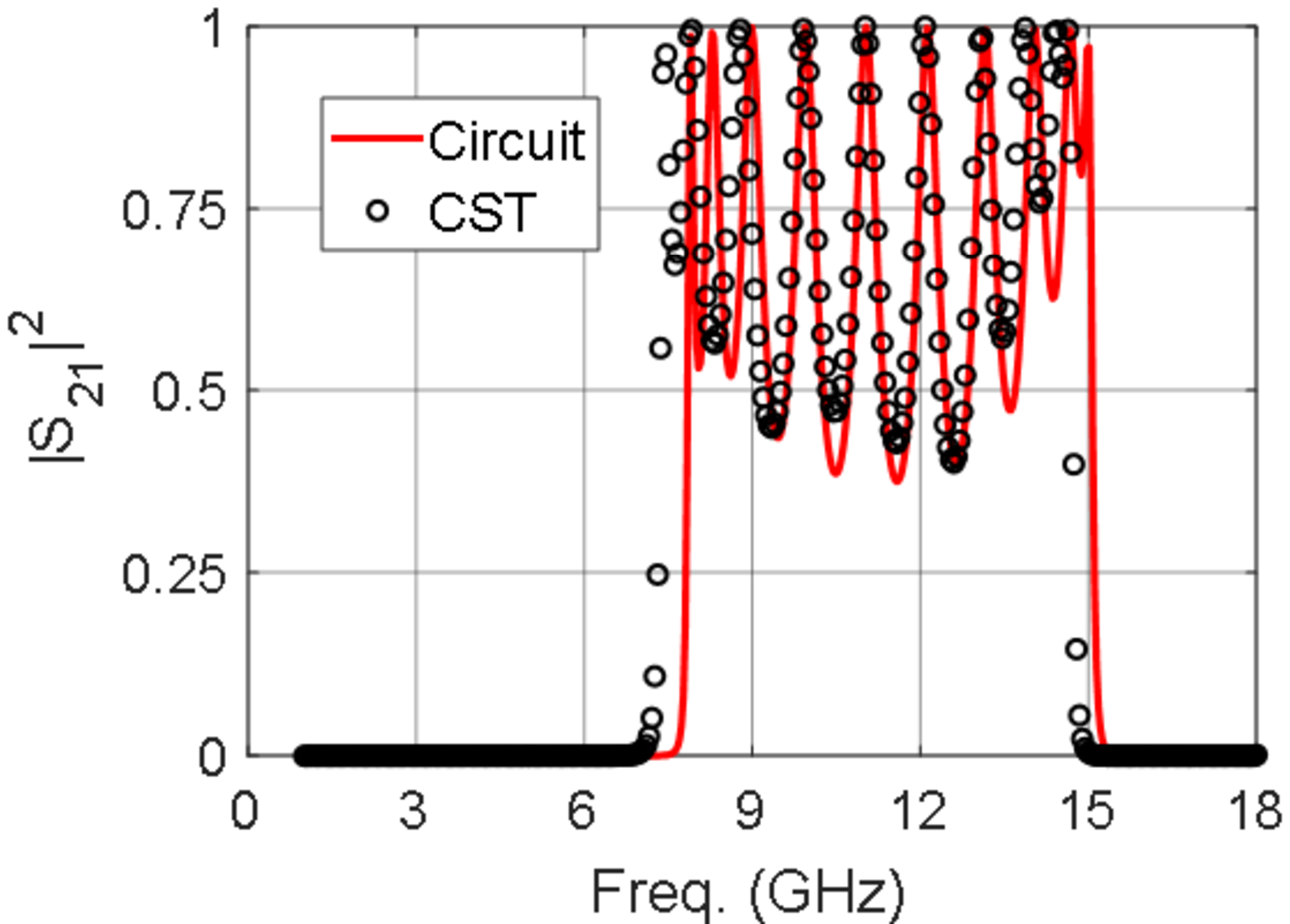}
	} 
	\caption{(a) Glide-symmetric stack of ten 2D arrays formed by subwavelength annular apertures. (b) Transmissivity versus frequency for TM normal incidence. Geometrical parameters of the unit cell: $a=3.8$\,mm, $b=4.8$\,mm, $p_x=p_y=p=10$\,mm,  $d_x=d_y=p/2$, $h=1.575$\,mm, and $\varepsilon_r=2.65$.  }
	\label{fig:10lay_glide}
\end{figure}

\Fig{fig:10lay_glide}(a) presents the glide-symmetric version of the stacked structure previously presented in \Fig{fig:10lay}(a), named henceforth as mirror-symmetric. For a fair comparison, the same geometrical parameters and number of layers have been kept. \Fig{fig:10lay_glide}(b) illustrates the transmissivity of the glide-symmetric staked structure. A good agreement is observed between the proposed formulation and the finite element method (FEM) of commercial software \textit{CST} for such a complex transmission spectra. However, our formulation is significantly more computationally efficient than the commercial software. Using the same computer, the ECA took less than 20~seconds in the analysis of the whole frequency range while \textit{CST} took more than 30~minutes for the same analysis.  It is also observed that the passband of the glide-symmetric configuration (8\,GHz of bandwidth) is notably widened compared to the mirror-symmetric structure (6\,GHz of bandwidth). This can be attributed to the suppression of the stopband related to the first Bloch mode in the glide-symmetric FSS. However, note that the ripple level of the structure with glide symmetry is also increased compared to the mirror-symmetric structure. This is associated with the mismatching of the impinging free-space wave and the impedance of the propagating Bloch mode in the stacked structure. These facts will be discussed in greater detail in Section~\ref{sec:3-D}.

\subsection{\label{sec:NonalignedB} Asymmetrical and Nonaligned Stacks}

\begin{figure}[h]
	\centering
	\subfigure[]{\includegraphics[width= 0.95\columnwidth]{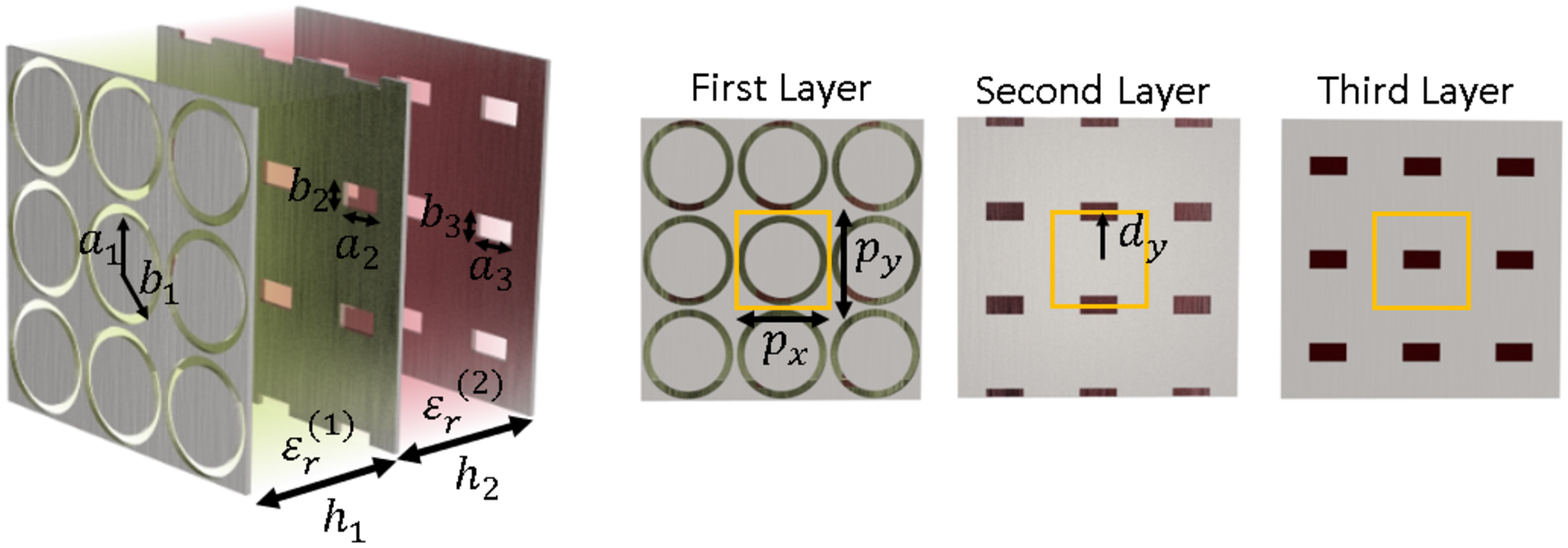}
	}\\
	\subfigure[]{\includegraphics[width= 0.8\columnwidth]{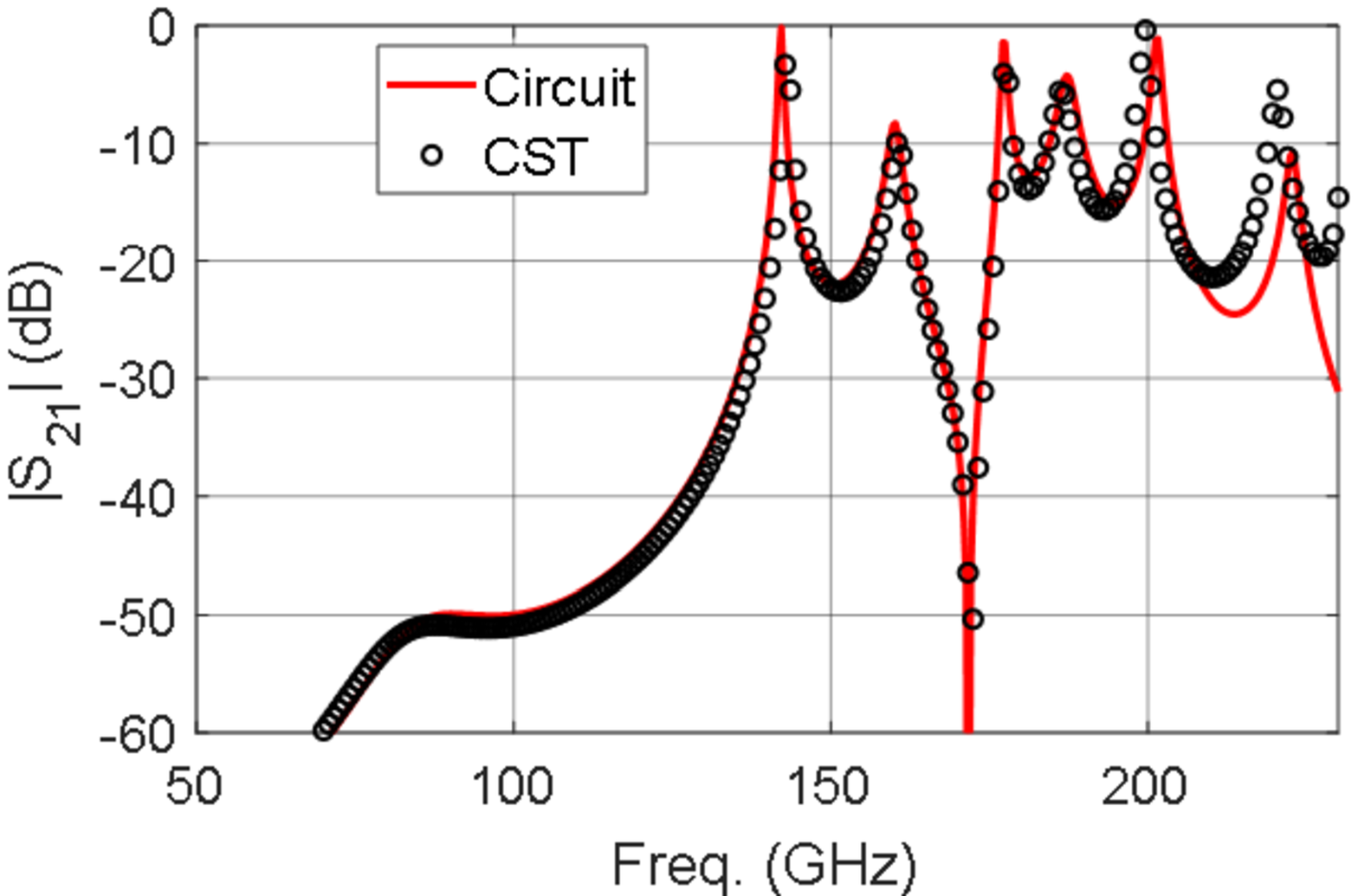}
	} 
	\caption{(a) Asymmetrical and nonaligned stack of two \mbox{2-D} arrays formed by annular and rectangular apertures. (b)~Magnitude of the transmission parameter for TM~normal incidence. Geometrical parameters of the unit cell:  $a_1=3.8$\,mm, $b_1=4.8$\,mm, $a_2=a_3=0.4$\,mm, $b_2=b_3=0.2$\,mm, $p_x=p_y=p=1$\,mm, $d_y=p/2$, $h_1=h_2=0.6$\,mm, and $\varepsilon_r^\mathrm{(1)}=\varepsilon_r^\mathrm{(2)}=2.6$. }
	\label{fig:annular_ring}
\end{figure}

The range of use of the circuit model is not limited to multilayered structures with the same type of apertures, such as those discussed above. Different types of apertures can be combined, as shown in~\Fig{fig:annular_ring}(a). Periodic annular and rectangular apertures are stacked in this case, forming a \mbox{3-layer} structure where the second perforated plate is off-shifted half a period in $y$~direction, $d_y=p/2$. The spatial profile of the electric field assumed to be excited on the rectangular aperture is given in~\eqref{spatialprofile_cos_sqrt} in the Appendix. Our closed-form results and the ones obtained with~\textit{CST} are plotted in~\Fig{fig:annular_ring}(b). Good agreement is observed with \textit{CST} in a wide frequency band, reaching a precision to the third and fourth decimal place (-60\,dB) in the rejection bands. It should be remarked the fully-analytical nature of our results, in contrast to previous approaches~\cite{arbitrary2D}. Fully-analytical results can be obtained as long as the spatial profile of the considered apertures can be expressed in a closed form, regardless of the geometry of the apertures and the application of linear transformations (displacement, rotation, scaling, etc).

\subsection{Rotated FSS}

As previously mentioned, rotation is one of the possible algebraic transformations that can relate the left and right apertures of a coupled pair. Stacks of rotated periodic structures is a practical configuration, usually applied in the context of polarization converters~\cite{Page-AP2018, Page-AP2020, Costa-AP2020}. In the frame of our analytical ECA, the spatial field profile of a single right- or left-side rotated aperture $\mathbf{E}_{\text{R/L}}^{\text{rot}}(\mathbf{r})$ admits to be represented in terms of the field profile in a non-rotated aperture $\mathbf{E}_{\text{L/R}}(\mathbf{r})$ through the rotation matrix $\underline{\mathbf{R}}$:
\begin{equation}\label{Erot}
    \mathbf{E}_{\text{a,R/L}}^{\text{rot}}(\mathbf{r}) = \underline{\mathbf{R}}\mathbf{E}_{\text{a,L/R}}( \underline{\mathbf{R}}^{-1}\mathbf{r})
\end{equation}
where $\mathbf{r}=x\hat{\mathbf{x}} + y\hat{\mathbf{y}}$ and
 \begin{equation}
\underline{\mathbf{R}} = 
\begin{bmatrix}
        \cos\alpha & -\sin\alpha\\
        \sin\alpha & \cos\alpha
\end{bmatrix}
\end{equation}   
with $\alpha$ being the rotation angle of the apertures in the counterclockwise direction. Starting from~\eqref{Erot}, it can be demonstrated that the \mbox{2-D} Fourier transform of the spatial profile in the rotated aperture can be written in terms of the profile with no rotation as~\cite{book-rotation,rotation}
\begin{equation}\label{FErot}
    \widetilde{\mathbf{E}}_{\text{a, R/L}}^{\text{rot}}(\mathbf{k}_{\text{t}, nm}) = \underline{\mathbf{R}}\widetilde{\mathbf{E}}_{\text{a, L/R}}(\underline{\mathbf{R}}^{-1}\mathbf{k}_{\text{t}, nm})\,. 
\end{equation}
The obtaining of the \mbox{2-D} Fourier transform of the rotated field-profile just demands a simple linear transformation in terms of the rotation angle~$\alpha$. The computation of the corresponding TM- and TE- transformer ratios is finally achieved by introducing  $\widetilde{\mathbf{E}}_{\text{a, R/L}}^{\text{rot}}(\mathbf{k}_{\text{t}, nm})$ in~\eqref{eq:NnmTM} and~\eqref{eq:NnmTE}.  

\begin{figure}[t]
	\centering
	\subfigure[]{\includegraphics[width= 0.9\columnwidth]{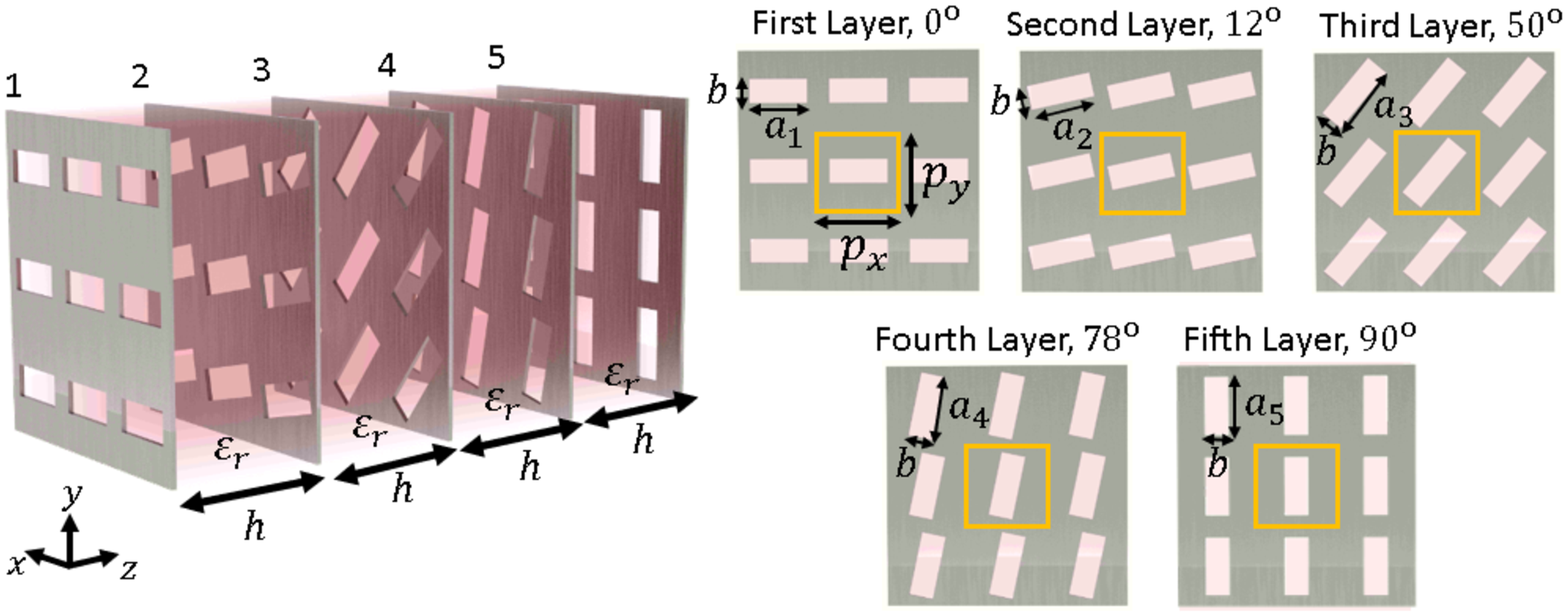}}
	\subfigure[]{\includegraphics[width= 0.7\columnwidth]{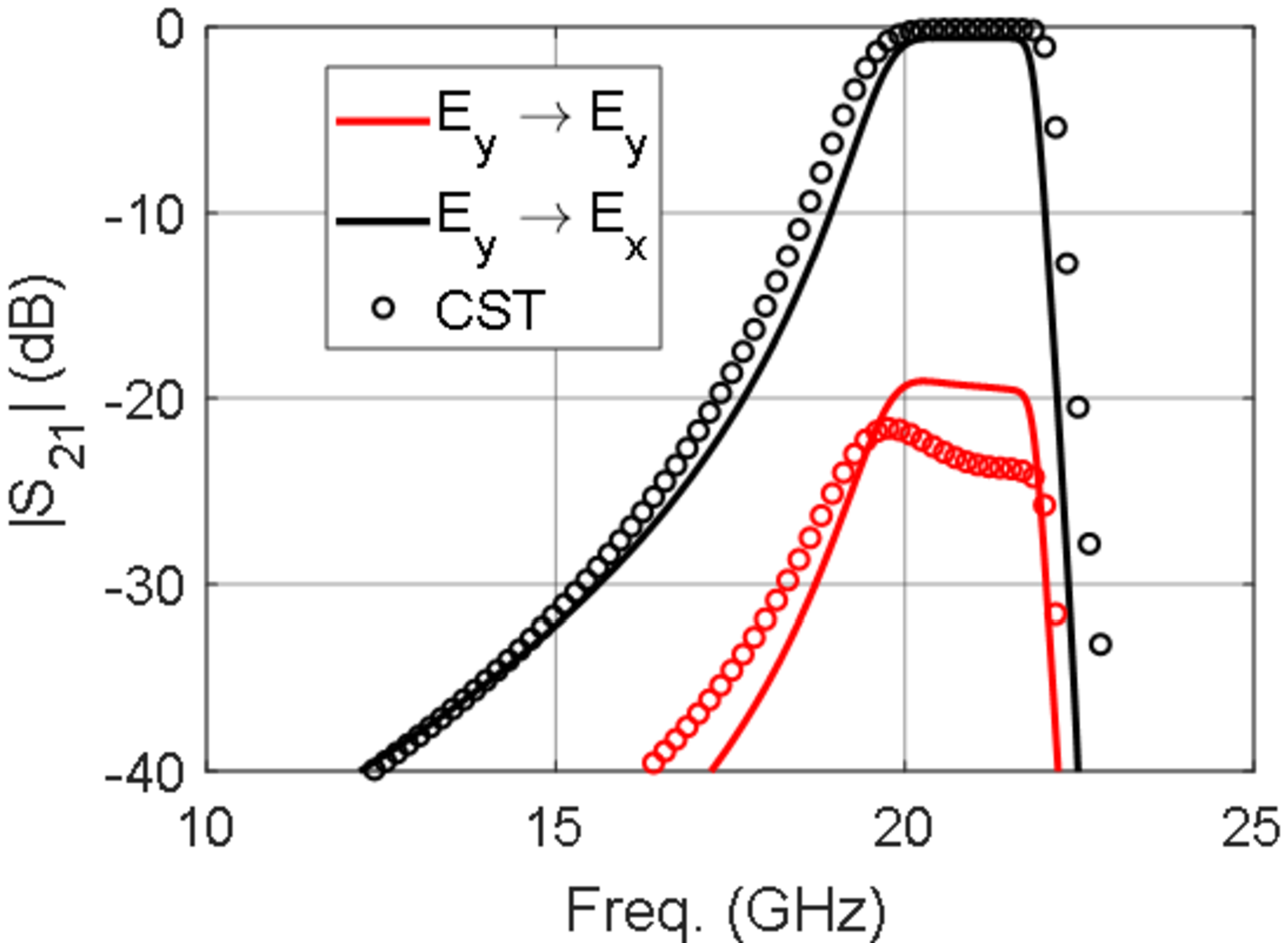}
	} 
	\caption{(a) Five-layer rotated stack formed by rectangular apertures. (b)~Magnitude of the transmission parameters for TM~normal incidence. Geometrical parameters of the unit cell: $a_1=a_5=7.25$\,mm, $a_2=a_4=8$\,mm, $a_3=9$\,mm, $b=3$\,mm, $p_x=p_y=p=10$\,mm, $h=1.5$\,mm, and $\varepsilon_r=1$.}
	\label{fig_rot}
\end{figure}

An example of a stack comprising five rotated free-standing FSSs ($q=1,2,\ldots,5$) is shown in~\Fig{fig_rot}(a), where it can be seen that each single metallic screen is periodically perforated with rectangular apertures of dimensions $a_q\!\times\!b_q$. The stack configuration is such that the rotation angle of the first and fifth (last) layers are $\alpha_{1} =0\degree$ and $\alpha_{5} = 90\degree$, respectively. This configuration, if efficiently optimized, may constitute a potential polarization converter. The orientation of the first and fifth apertures is suitable for a conversion from $y$-polarized to $x$-polarized electric fields. To accomplish the conversion, the intermediate layers ($q = 2, 3, 4$) have to be conveniently designed. In the example of~\Fig{fig_rot}(a), the corresponding rotation angles are $\alpha_{2} = 12\degree$, $\alpha_{3} = 50\degree$, and $\alpha_{4} = 78\degree$, following a growing trend from the first to the fifth screen. Along with the optimum rotation angle for each screen, the dimensions of each aperture are also optimized. In particular, an optimum configuration has been found by keeping the shorter dimension of the apertures~($b_q$) identical for all the screens whereas the larger dimension~($a_{q}$) is symmetrically distributed ($a_{1} = a_{5}$, $a_{2} = a_{4}$). All the above derivations could be carried out with a very reduced computational effort due to the analytical nature of the employed ECA.  

\Fig{fig_rot}(b) shows the magnitude of the transmission coefficient of the above structure when it is excited by a normally-impinging plane wave with the electric-field vector directed along the~$\hat{\mathbf{y}}$~direction. The transmission coefficient is split into two components: the component corresponding the co-polarization term ($E_y\!\to\! E_y$) and the component corresponding to the cross-polarization term ($E_y\!\to\! E_x$). As it can be appreciated in the figure, almost full conversion $E_y\!\to\! E_x$ is achieved from~$20$ to~$22$\,GHz, covering a  fractional bandwidth of $9.5\%$ approximately (the co-pol level is below -20\,dB). The agreement between results provided by \textit{CST} and the results obtained by our analytical circuit model is very good. As a comparison of the required computational effort, \textit{CST} took more than fifteen minutes (900~seconds) to compute 1001 equally-spaced frequency points while the proposed ECA took less than 15~seconds.

\subsection{\label{sec:metal-backed} Metal-backed FSS}
The present ECA can easily deal with scenarios where the stack of aperture arrays are backed by a metallic screen. The back metallic screen is simply modeled as a short circuit; namely, the transmission lines associated with harmonics in direct contact with the ground plane have to be terminated with a short circuit. In the circuit representation shown in~\Fig{fig:coupled}, the outgoing dielectric medium was assumed semi-infinite and has an equivalent admittance, $Y_\text{ext}^{\text{out}}$, coming from the infinite transmission lines in parallel shown in the yellow box of the equivalent circuit. When this medium is grounded, the equivalent admittance $Y_{\text{ext}}^{(\text{out})}$ defined in~\eqref{eq:Yext} now becomes
 \begin{multline}\label{eq:Yext_short}
    Y_\text{ext}^\text{(out)} = \sum_{\substack{n,m=-\infty}}^\infty \!\!  \Big[- \jj \left(N_{nm,\text{R}}^{(2),\text{TM}}\right)^2  \,Y_{nm}^{(\text{out}),\text{TM}} \cot(k_{z, nm}^{(\text{out})}h_{\text{out}}) \\ - \jj
    \left(N_{nm,\text{R}}^{(2),\text{TE}} \right)^{2} \,Y_{nm}^{(\text{out}),\text{TE}} \cot(k_{z,nm}^{(\text{out})}h_{\text{out}}) \Big]
\end{multline}
where $Y_{nm}^{(\text{out}),\text{TM/TE}}$ and $k_{z,nm}^{(\text{out})}$ take into account the relative permittivity of the grounded dielectric medium, and~$h_{\text{out}}$  is the length of the shorted transmission lines (that is, the length of the grounded dielectric medium).  Unlike~\eqref{eq:Yext}, the summation in \eqref{eq:Yext_short} has now to include the fundamental harmonics of order $n = m = 0$ for both TE and TM harmonics.

\begin{figure}[t]
	\centering
	\subfigure[]{\includegraphics[width= 0.95\columnwidth]{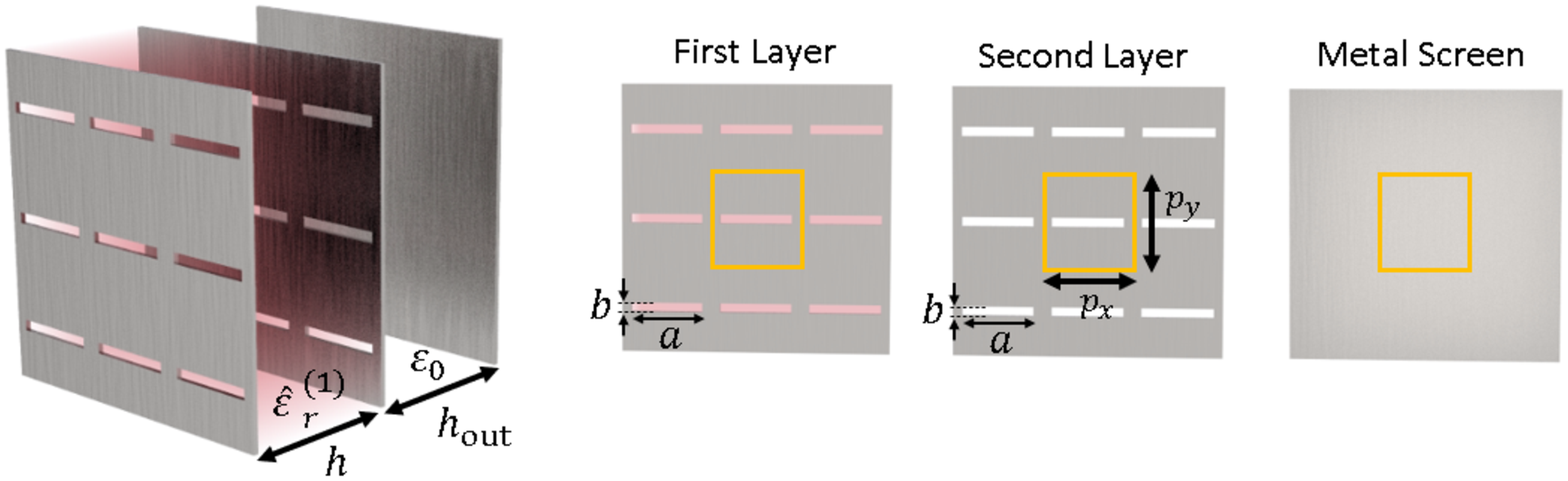}
	}\\
	\subfigure[]{\includegraphics[width= 0.49\columnwidth]{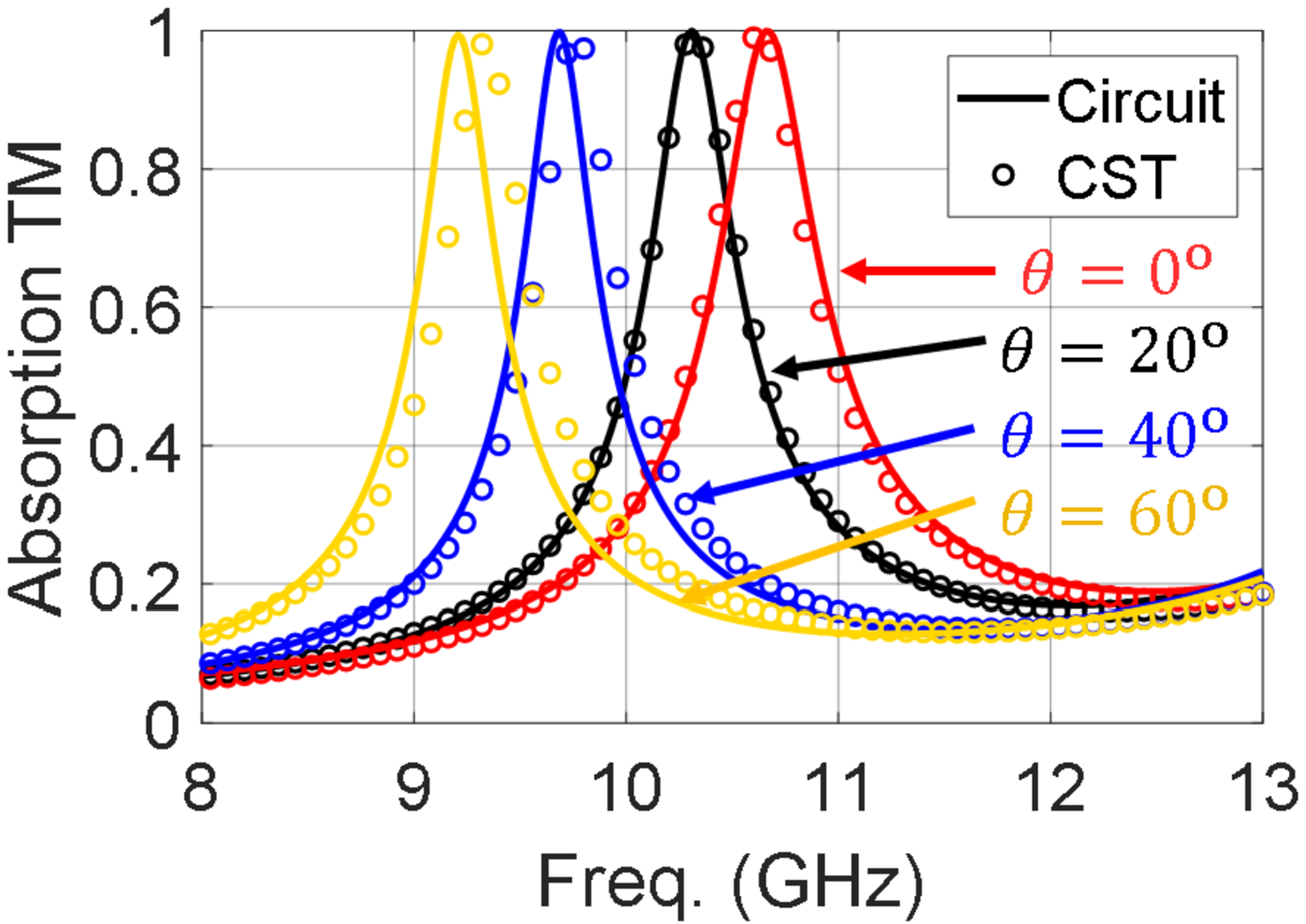}
	}
	\hspace*{-0.4cm}
	\subfigure[]{\includegraphics[width= 0.49\columnwidth]{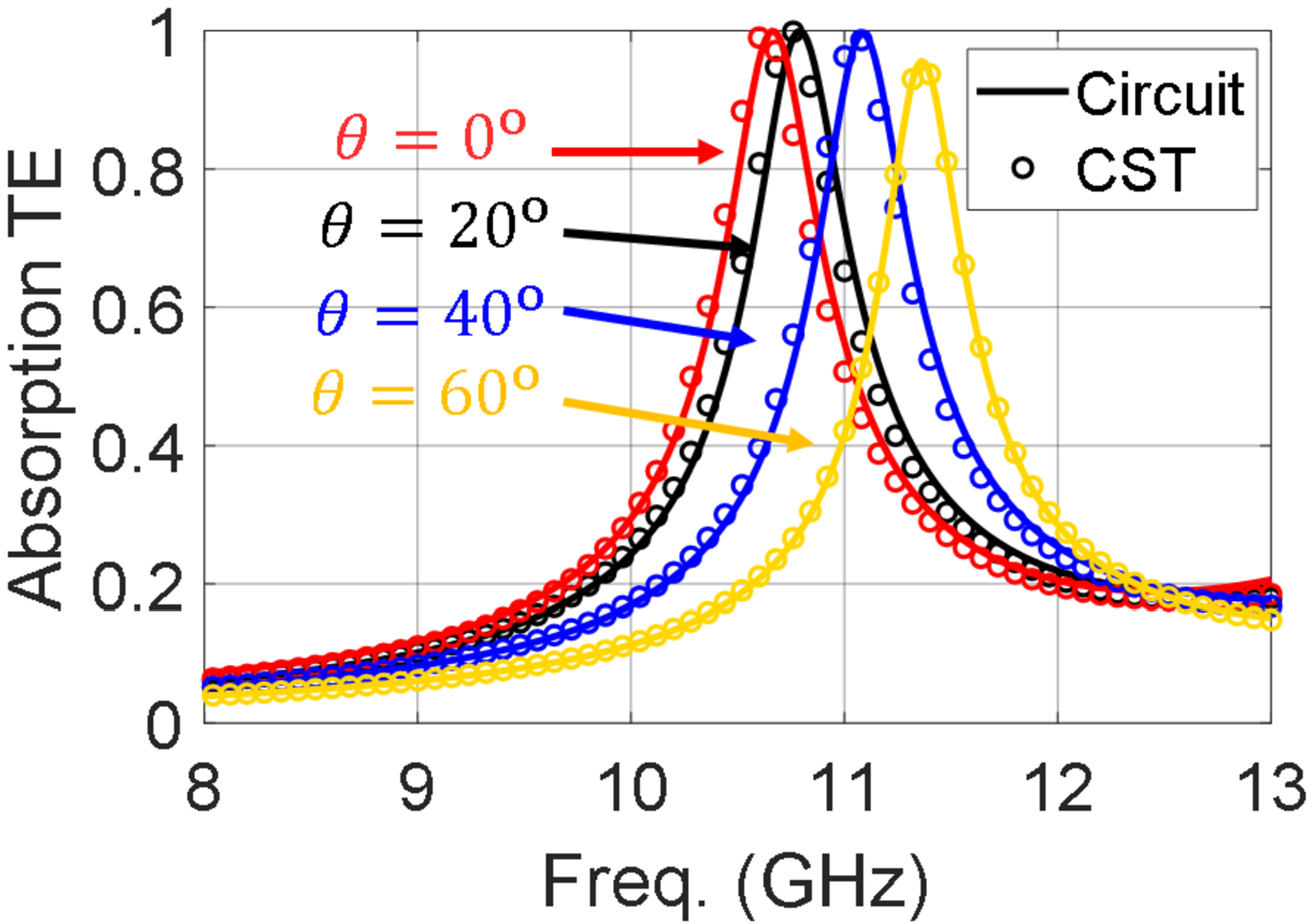}
	} 
	\caption{(a) Metal-backed stacked structure acting as an absorber. Absorption coefficient for (b)~TM~oblique incidence and (c)~TE~oblique incidence. Geometrical parameters of the unit cell:  $a=8$\,mm, $b=1 $\,mm,  $p_x=p_y=p= 10$\,mm,  $h=0.75$\,mm, $h_{\text{out}} = 0.75$\,mm, and $\hat{\varepsilon}_{\text{r}}^{(1)} = 4\times(1 - \jj 0.02)$.} 
	\label{fig:absorber}
\end{figure}

Grounded dielectric FSSs can be employed for the design of absorbers by introducing, for instance, a lossy dielectric substrate in the stack. The information about the losses is included in the permittivity of the lossy material, which thus becomes a complex quantity. An example of a possible absorber configuration is depicted in~\Fig{fig:absorber}(a), where a stack of two perforated metallic screens with identical rectangular apertures is considered. A lossy dielectric slab of FR4 is sandwiched between both layers, having a relative permittivity of $\varepsilon_{r}^{(1)} = 4$ and loss tangent $\tan(\delta) = 0.02$ ---the complex permittivity of the lossy substrate is then given by $\hat{\varepsilon}_{\text{r}}^{(1)} = 4\times(1 - \jj 0.02)$. A free-standing ground plane ($\varepsilon_r^\text{(out)}=1$) is placed at a distance~$h_{\text{out}}$. The structure is assumed to be excited by either a TM-polarized or a TE-polarized plane wave that impinges obliquely with an angle~$\theta$ (normally incident waves are considered when $\theta=0\degree$). The geometry of the unit-cell aperture is chosen to have a high absorption rate in both polarizations up to a incidence angle of 20 degrees, though in a narrow frequency band as shown in~Fig.\,\ref{fig:absorber}(b) and~(c). 

The absorption coefficient ($A_c$) is calculated in terms of the reflection coefficient as $A_\text{c} = \sqrt{1 - |S_{11}|^{2}}$,
where it is assumed that the reflected power is only carried by the $n,m=(0,0)$ harmonic, whose reflection coefficient is then given by~$S_{11}$. This assumption is valid below the onset of grating lobes and absence of cross-pol effects, which is fully satisfied in the present case (the symmetry of the rectangular apertures prevents the excitation of the cross-polarization term).  The strong coupling induced by the proximity of the perforated plates can result in high absorption amplitudes, as this case illustrates. The thickness of the whole absorber is very reduced, with a total size of $1.5$\,mm, which means a thickness of $\lesssim \lambda_0/20$ at the full-absorption peaks observed in the figures. A slight frequency shift of less than~1\% can be appreciated between some results given by \textit{CST} and the ECA. The obtained good accuracy of the ECA demonstrates the ability of the approach to cover scenarios with stacked arrays including the presence of dielectric losses and a reflecting ground plane. The ability of the proposed ECA to deal with oblique incident waves is also validated in this example. Scenarios with $\theta=60^{o}$ are well represented for both TM and TE incidence. 

It should be noted that the present high accuracy of the ECA is directly related to the good approximation that the assumption made in~\eqref{eq:Et} stands for in this case. The accuracy is expected to be more limited in other situations where the eventual excitation of higher resonances of the aperture is more significant. The lack of symmetry in the problem, the electrical size of the aperture as well as the number of plates are factors that may affect the suitability of the implicit approximation of the ECA in scenarios where oblique incidence is considered. However, it should be pointed out that many practical situations do satisfy the ECA validity conditions, and that is the model's application niche this work is exploring.

\section{\label{sec:Arbitrary} Arbitrary Apertures}
For non-canonical aperture geometries for which the spatial profile in~\eqref{eq:Et}  is not easily expressed in closed form, we can make use of a hybrid approach that advantageously combines the use of the equivalent circuit and commercial simulators~\cite{patches2}. More specifically, we can benefit of the ability of commercial simulators to deal with arbitrary geometries to extract the spatial profile of the considered aperture when only a \textit{single} periodic array is considered in free space. The extraction of this spatial profile has to be done at a \textit{single} frequency point, which is an operation far less computationally demanding than the simulation of the complete stack in the full frequency range. From this operation, the transformer turn ratios in~\eqref{eq:NnmTM} and~\eqref{eq:NnmTE} are obtained after numerically computing  the Fourier transform of the aperture spatial profile. Moreover, linear transformations (rotation, scaling, displacement) can still be applied to this Fourier transform in order to find relations between the apertures of the perforated plates. As previously discussed, this approach will be valid as long as the aperture spatial profile does not greatly vary with frequency.

\begin{figure}[t]
	\centering
	\subfigure[]{\includegraphics[width= 1\columnwidth]{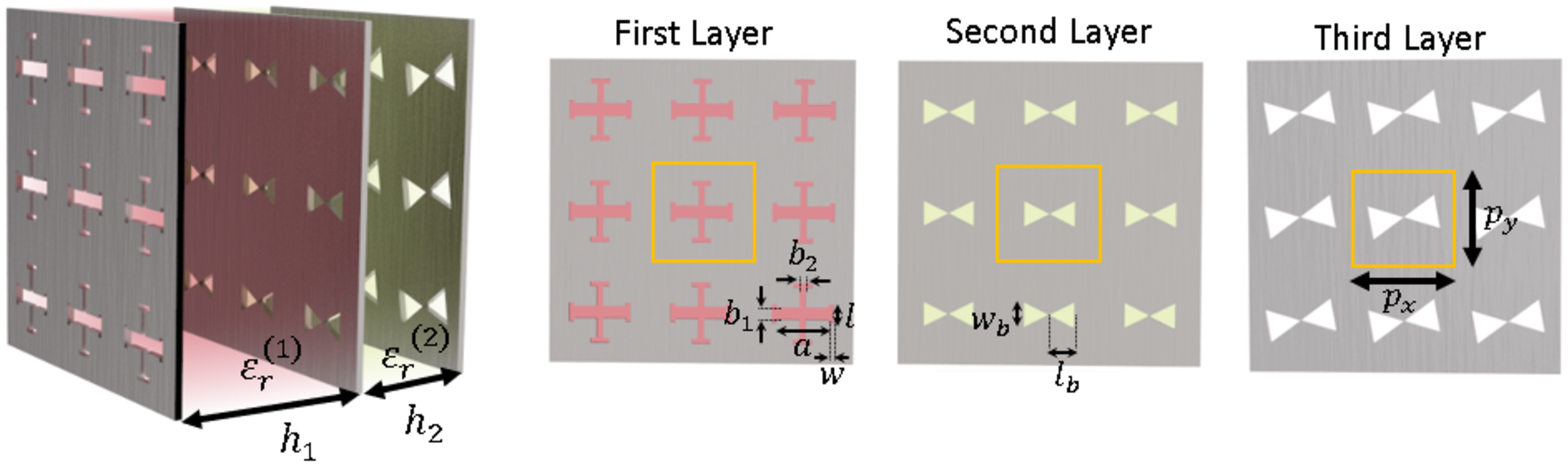}}
	\subfigure[]{\includegraphics[width= 0.9\columnwidth]{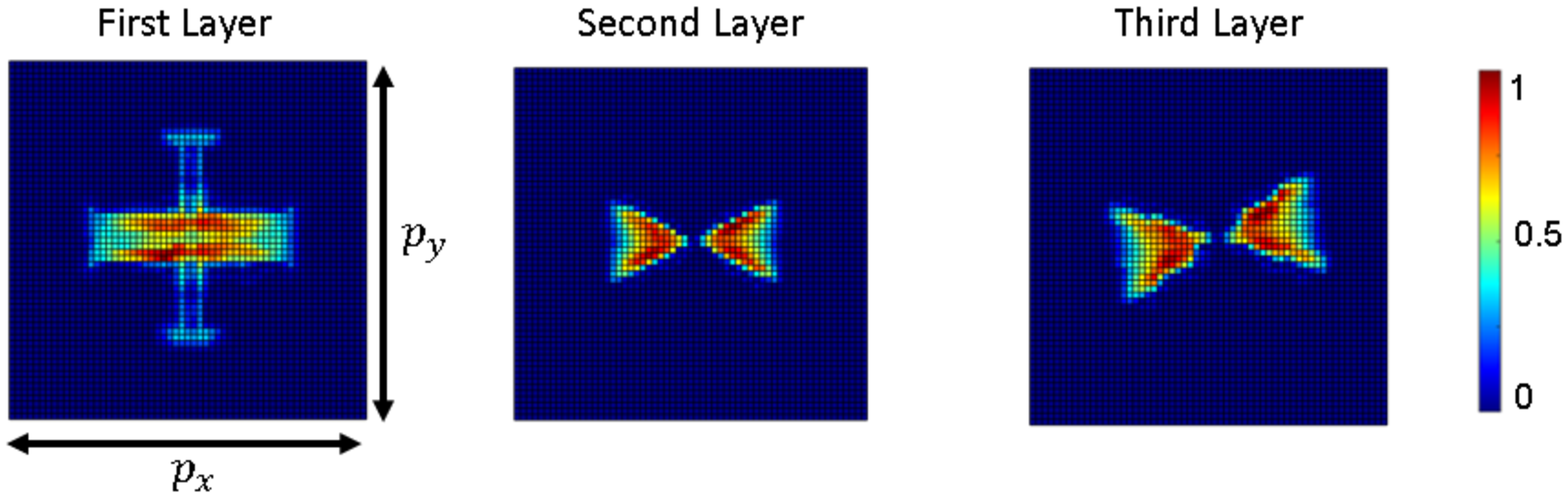}}
	\subfigure[]{\includegraphics[width= 0.85\columnwidth]{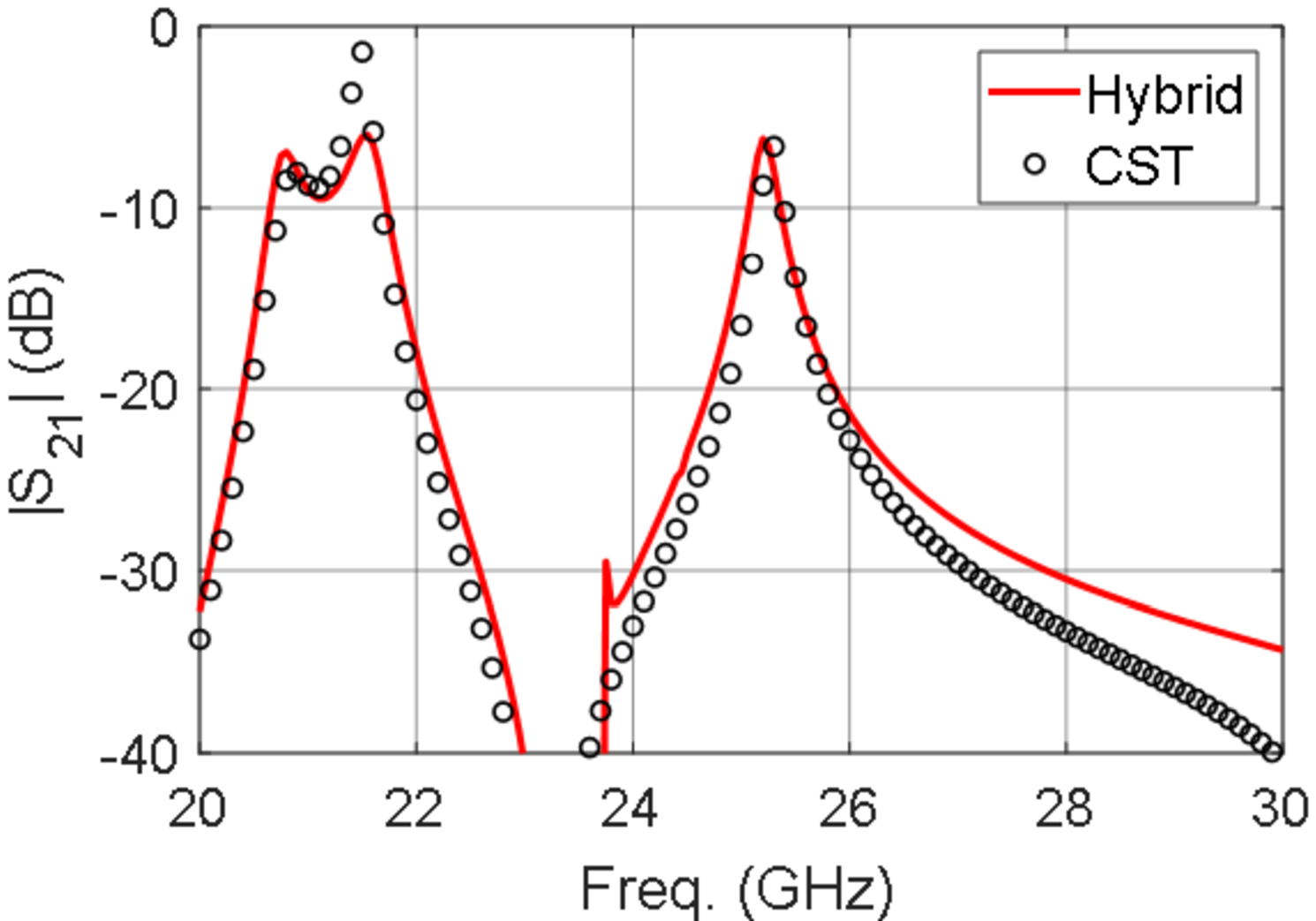}}
	\caption{Three-layer stack of arbitrary apertures. Geometrical parameters of the unit cell: $a=4.1$\,mm, $b_1=1$\,mm, $b_2=0.5$\,mm, $l=1.5$\,mm, $w=0.4$\,mm, $l_{b}=2$\,mm, $w_{b}=2$\,mm, $p_x=p_y=p=8$\,mm, $h_1=1.575$\,mm, $h_2=0.5$\,mm, $\varepsilon_r^\mathrm{(1)}=4.7$, and $\varepsilon_r^\mathrm{(2)}=2.5$. The bowtie-shaped apertures of the third layer are rotated $\alpha=10^\mathrm{o}$ and scaled 1.2 times ($S_x=S_y=1.2$) with respect to the apertures of the second layer.}
	\label{fig:arb}
\end{figure}

As an example, \Fig{fig:arb}(a) presents a 3-layer stack formed by a periodic array of Jerusalem-cross apertures and two arrays of bowtie-shaped apertures, separated by two different dielectrics of permittivities, $\varepsilon_r^\mathrm{(1)}=4.7$ and $\varepsilon_r\mathrm{(2)}=2.5$. The spatial profiles of the different apertures are extracted  with the simulation of single, free-standing layers in~\textit{CST} at the lowest operating frequency (20\,GHz). In this example, the mesh of the unit cell consists of 60$\times$60 hexahedral elements. In order to show the potentialities of the approach also for arbitrary geometries, the apertures of the third layer are taken as rotated and scaled versions of the apertures that form the second layer. This can be appreciated in~\Fig{fig:arb}(b), where the absolute value of the aperture spatial profiles is shown.  Thus, in this case it is only necessary to extract the spatial profile of the first and second plates $\mathbf{E}_{a,\text{L}}^{(1,2)}(x,y)$, since the Fourier transform of the spatial profile of the third plate can be expressed as $\widetilde{\mathbf{E}}_{a,\text{R}}^{(2)}(\mathbf{k}_{t,nm})= \underline{\mathbf{R}} \, \underline{\mathbf{S}} \, \widetilde{\mathbf{E}}_{a,\text{L}}^{(2)}(\underline{\mathbf{R}}^{-1} \underline{\mathbf{S}}^{-1} \mathbf{k}_{t,nm})$, where $\underline{\mathbf{R}}$ is the previously defined rotation matrix and  $\underline{\mathbf{S}}$ represents a scale matrix given by \cite{book-rotation, rotation}
\begin{equation}
\underline{\mathbf{S}} = 
\begin{bmatrix}
        S_x & 0\\
        0 & S_y
\end{bmatrix}\;.
\end{equation}
Note that matrices $\underline{\mathbf{R}}$ and $\underline{\mathbf{S}}$ commute if $S_x=S_y$, as the scaling operator is then defined by a diagonal matrix. \Fig{fig:arb}(c) illustrates the transmission parameter of the 3-layer stack of arbitrary apertures. A good agreement is observed with \textit{CST}, even for such a complex structure. Slight differences are observed beyond 26 GHz. This is due to the excitation of the second resonance in the Jerusalem cross, which reduces the range of validity of the ECA as the assumption of a frequency-independent spatial profile in \eqref{eq:Et} is no longer valid from this frequency. As a comparison, \textit{CST} took more than 20 minutes (more than 1200\,seconds) to compute 501 equally-spaced frequency points while the present hybrid approach took, in total, less than 40 seconds. It clearly proves that efficiency of the present hybrid implementation to compute the scattering parameters of stacked structures with arbitrary apertures.

\section{\label{sec:3-D} Infinite Periodic Stacks}

\subsection{Dispersion Diagram}
As is well known, the dispersion relation of a periodic structure along the $z$~direction can be expressed in terms of the elements of the transfer matrix of the corresponding unit cell (period $p_z\equiv p$) as~\cite{Collin, pozar}
\begin{equation} \label{eq:disp1}
    \cosh \left( \gamma_z p_z \right) = \frac{A_p + D_p}{2}
\end{equation}
where $\gamma=\alpha_z+\jj\beta_z$ is the propagation constant of the Floquet mode and the subindex $p$ means that $A_{p}$ and $D_p$ are elements of the transfer (ABCD) matrix $\underline{\mathbf{T}}_p$ associated with the unit cell of longitudinal
period~$p_z$. 
Given that $A=-Y_{22}/Y_{21}$, $D=-Y_{11}/Y_{21}$~\cite{pozar}, the dispersion relation can be expressed in closed form by replacing \eqref{eq:Yuv}--\eqref{Y22nm} into~\eqref{eq:disp1} to give 
\begin{multline} \label{eq:disp2}
    \cosh \left( \gamma_z p_z \right) \\ 
    = \dfrac{\sum\limits_{n,m=-\infty}^\infty \!\! \left[ \left({N_{nm,\text{L}}^{\text{TM/TE}}}\right)^2 + \left({N_{nm,\text{R}}^{\text{TM/TE}}}\right)^2 \right] 
    Y_{nm}^{\text{TM/TE}}\cot(k_{z,nm}h)}
    {2\sum\limits_{n,m=-\infty}^\infty \!\! N_{nm,\text{R}}^{\text{TM/TE}}\,{N_{nm,\text{L}}^{\text{TM/TE}}}\,
      Y_{nm}^{\text{TM/TE}} \csc(k_{z,nm}h) }
\end{multline}
where the sum above extends to both TE and TM modes. The index $(i)$ is removed in the above expression since all the internal regions are exactly the same in the present infinitely periodic structure under study. If there are more than one dielectric between the pair of coupled arrays, the terms inside the brackets in the second line of the above equation should be appropriately modified. In many practical situations, the unit cell of the periodic structure can be chosen to be symmetric, which implies that $A_p=D_p$ or, equivalently, ${N_{nm,\text{R}}^{\text{TM/TE}}} = {N_{nm,\text{L}}^{\text{TM/TE}}}$. 

As already mentioned in Sec.\,\ref{sec:cons}, one of the main limitations of the ECA for the study of the scattering properties of stacked  structures comes from the inadequacy of the method to deal with frequency sweepings where the spatial profile of the tangential field on the apertures significantly varies along the considered frequency range~\cite{eca_magazine}. This fact was physically linked to the excitation of higher resonances in the aperture.  From a practical point of view, this limitation can be overcome by expressing the spatial profile with more than one basis functions, as exploited, for instance, in~\cite{patches3}. However, at the light of~\eqref{eq:disp2}, one can observe that this equation can still be utilized to compute the dispersion behavior of high-order Floquet modes, despite using a single basis function to model the spatial profile on the aperture, as long as the single spatial profile employed to compute the transformer ratios in~\eqref{eq:disp2} can match the geometrical variations of the corresponding Floquet mode. In practice, it means that the function $\mathbf{E}_a(x,y)$ should be chosen with a spatial profile that closely resembles the different resonant modes of a single aperture. 

\begin{figure}[t]
	\centering
	\subfigure[]{\includegraphics[width= 0.88\columnwidth]{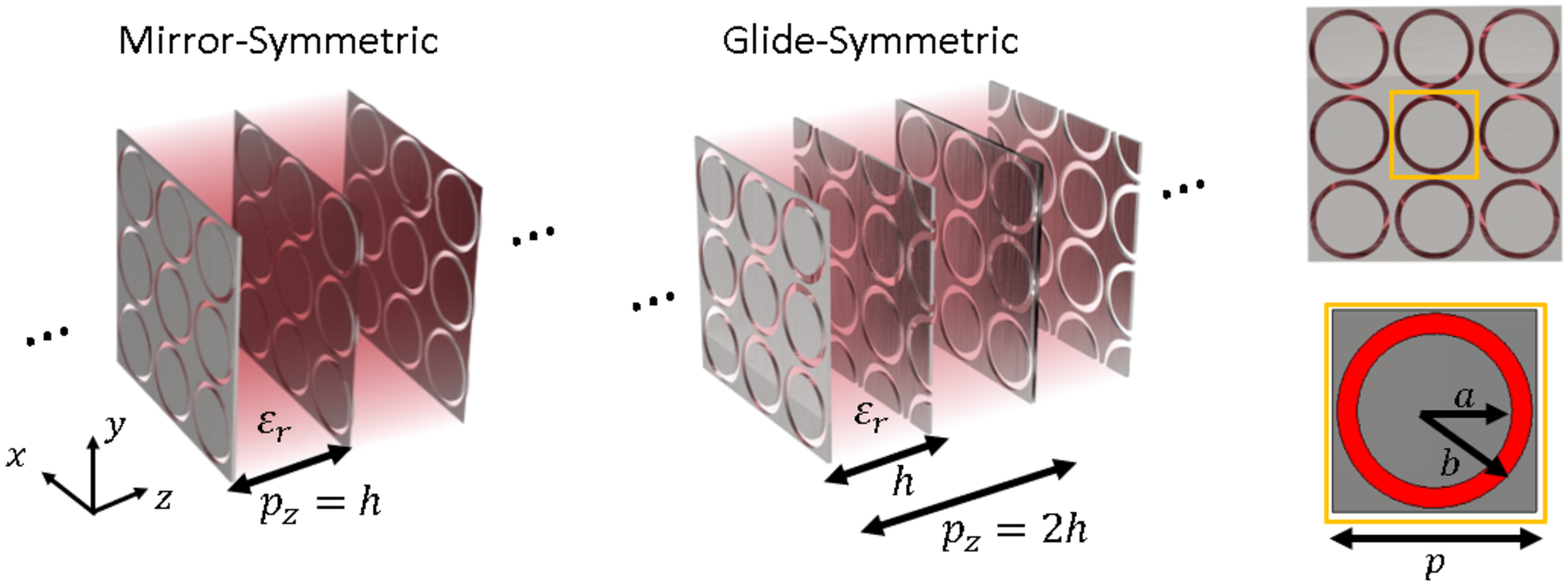}
	} 
	\subfigure[]{\includegraphics[width= 1\columnwidth]{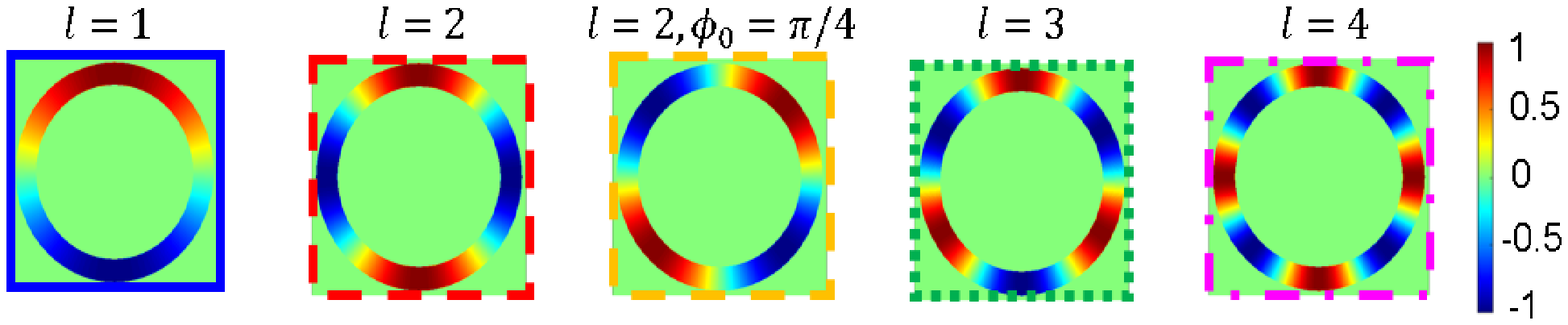}
	} 
	\subfigure[]{\includegraphics[width= 1\columnwidth]{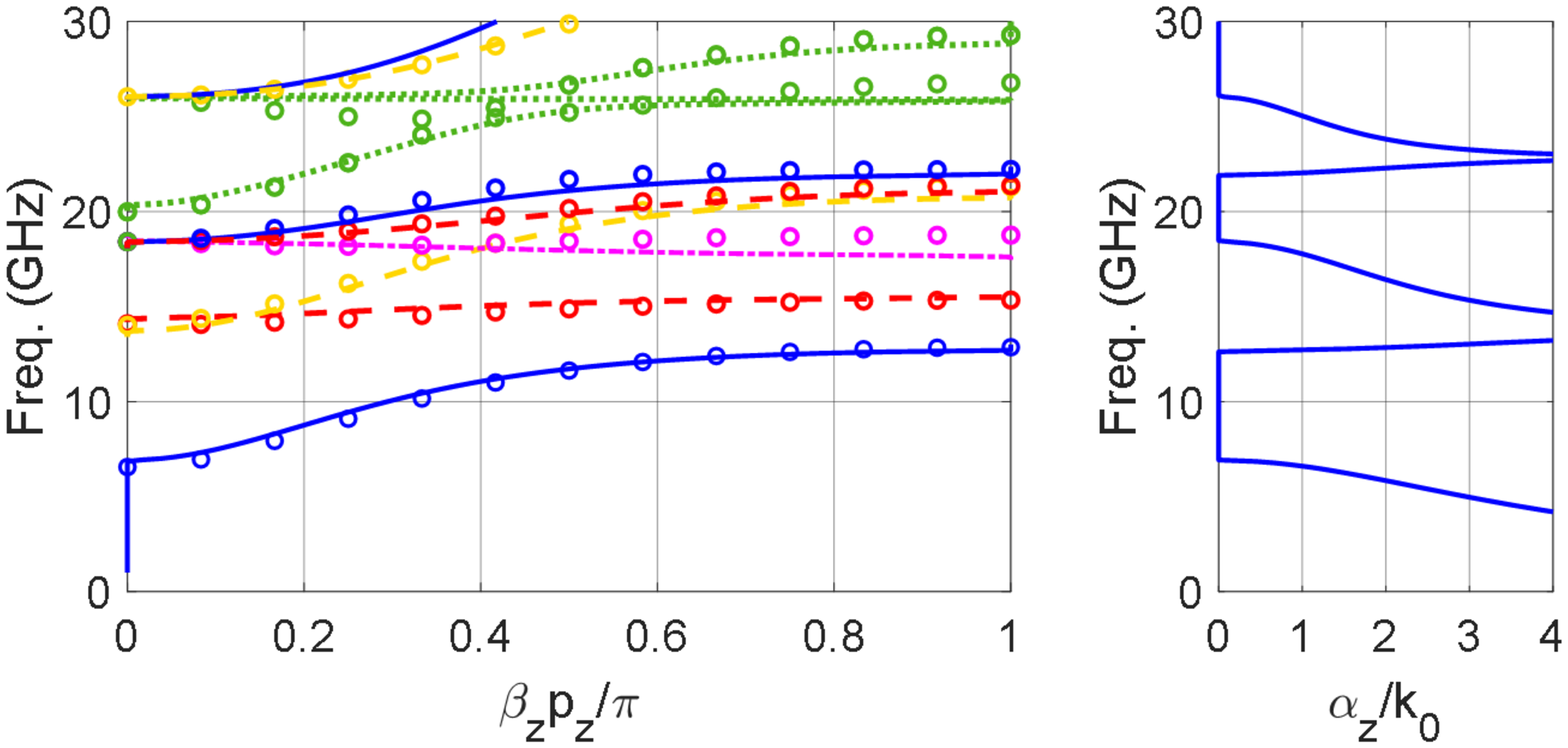}
	} 
	\subfigure[]{\includegraphics[width= 1\columnwidth]{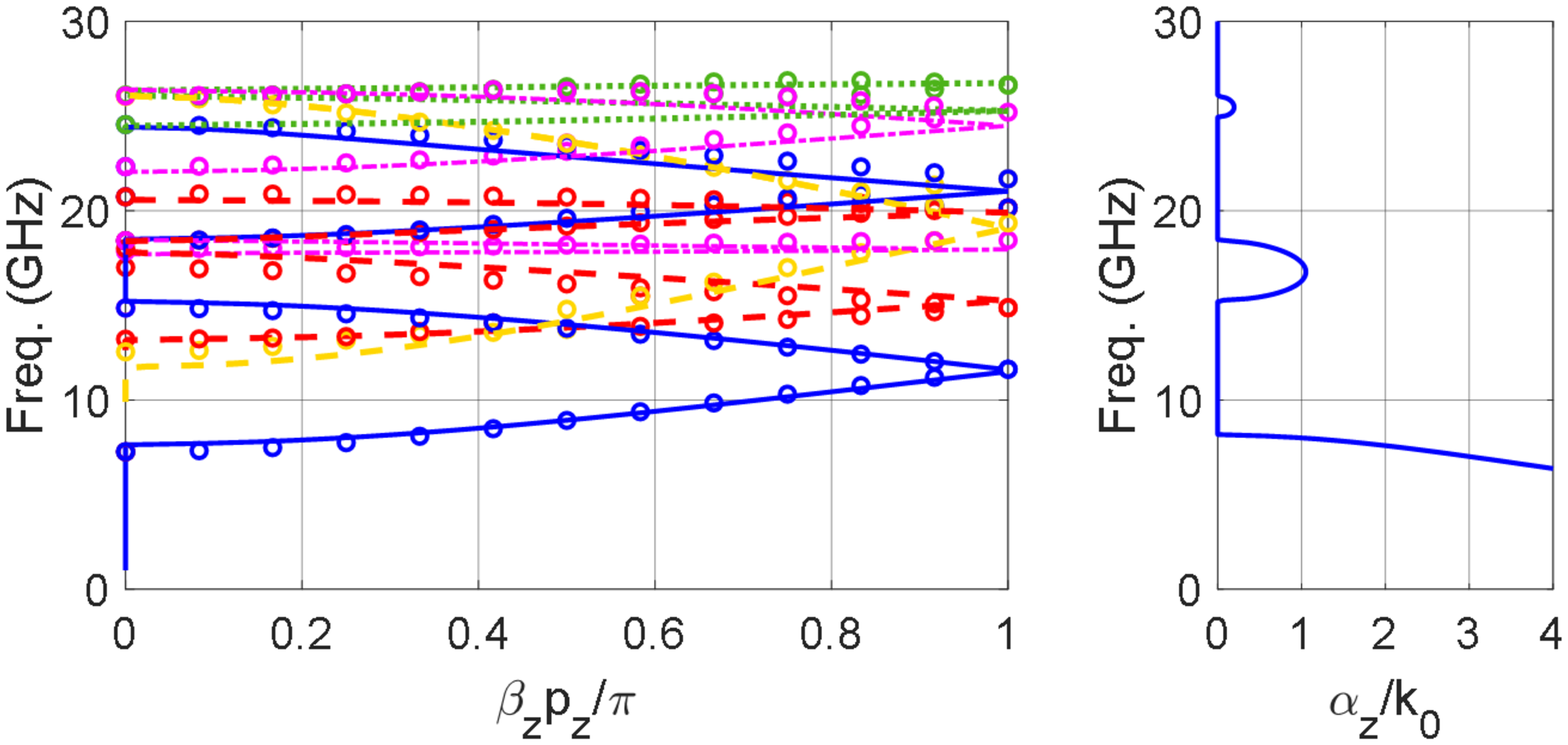}
	} 
		\caption{(a) Infinite periodic 3D array formed by annular apertures. (b)~Spatial distributions on the annular apertures analytically computed with Eq.\,\eqref{eq:Ea_anular}. Dispersion diagrams (phase shift and normalized attenuation constant) of the (c)~mirror-symmetric and (d)~glide-symmetric configurations.  The results extracted from \textit{CST} (colored circles) are shown for comparison purposes.  Parameters of the unit cell: $a=3.8$\,mm, $b=4.8$\,mm, $p_x=p_y=p=10$\,mm,  $p_z=1.575$\,mm, and $\varepsilon_r=2.65$.}
	\label{fig:disp}
\end{figure}

Following this rationale, the dispersion relation of the periodic mirror-symmetric stack of annular-aperture arrays shown in~\Fig{fig:disp}(a) has been computed by solving
\begin{equation} \label{eq:disp-mirror}  
     \cosh \left( \gamma_z h \right) =  \dfrac{\sum\limits_{n,m=-\infty}^\infty \!\!\!\!  \left(N_{nm}^{\text{TM/TE}}\right)^2 \left[  Y_{nm}^{\text{TM/TE}}\cot(k_{z,nm}h)\right]  }%
     {\sum\limits_{n,m=-\infty}^\infty \!\!\!\! \left(N_{nm}^{\text{TM/TE}}\right)^2 
     \left[  Y_{nm}^{\text{TM/TE}} \csc(k_{z,nm}h)\right] }\, .
\end{equation}
The indexes L/R has been suppressed since its difference is not necessary in this case. he different Floquet modes supported by the periodic stack are computed with \eqref{eq:disp-mirror} by imposing the spatial profiles shown in~\Fig{fig:disp}(b). In the case of annular apertures, these spatial profiles correspond to the mathematical form given in~Eq.\,\eqref{eq:Ea_anular} in the Appendix for $l=1,2,3,4$ and $\phi=0,\pi/4$. These closed-form expressions of the spatial profiles associated with the Floquet modes are found to match quite well the actual tangential fields extracted with \textit{CST}. A comparison of the results for the phase shift ($\beta_zp_z/\pi$) obtained with the simplified procedure proposed in this work and data provided by the \textit{CST} Eigensolver is shown in the left plot of~\Fig{fig:disp}(c), showing a good agreement between both sets of results.

The dispersion diagram of the glide-symmetric version of the above periodic structure is shown in~\Fig{fig:disp}(d). For glide-symmetric periodic structures, it should be taken into account that the actual unit cell  of the structure is symmetric (and of size $p_z=2h$ in the particular case under study); namely, \eqref{eq:disp1} reduces to 
\begin{align}\label{eq:disp12}
    \cosh \left( \gamma_z p_z \right)  & = A_p \;.
\end{align}
Since the actual unit cell now involves two pair of coupled arrays (and in the simplest case considered here, two dielectric layers of size $h$), it means that $A_{p}$ actually comes from $\underline{\mathbf{T}}_{p} = \underline{\mathbf{T}}'_{p/2} \underline{\mathbf{T}}''_{p/2}$, where $\underline{\mathbf{T}}'_{p/2}$ stands for the transfer matrix of one of the two sub-unit cells of size $p/2$ that comprises the actual unit cell ($\underline{\mathbf{T}}''_{p/2}$ is the transfer matrix of the remaining sub-unit cell). In similarity with the discussion in~\cite[Sec.\,2.2]{MesaSymmetry08}, the original dispersion relation~\eqref{eq:disp1} of the glide-symmetric structure [with period $p_z=2h$, as shown in~\Fig{fig:disp}(a)] can alternatively be rewritten as
\begin{equation} \label{eq:disp-glide}
    \cosh \left( \gamma_z p_z/2 \right) = \sqrt{A_{p/2}D_{p/2}} = A_{p/2}\;. 
\end{equation}
For glide-symmetric structures, $A_{p/2}$ turns out to be equal to $D_{p/2}$, which  follows after introducing~\eqref{eq:NRNL_glide} into~\eqref{Y11nm} and~\eqref{Y22nm}.
It implies that the dispersion relation of glide-symmetric structures can be obtained dealing only with the sub-unit cell of the structure (of size~$h$ in the present case); namely, the dispersion equation can be written as
\begin{multline} \label{eq:disp-glide1}  
     \cosh \left( \gamma_z h \right) \\
     =  \dfrac{\sum\limits_{n,m=-\infty}^\infty \!\!\!\! \left(N_{nm}^{\text{TM/TE}}\right)^2 \left[  Y_{nm}^{\text{TM/TE}}\cot(k_{z,nm}h)\right]  }%
     {\sum\limits_{n,m=-\infty}^\infty \!\!\!\! (-1)^{n+m}\left(N_{nm}^{\text{TM/TE}}\right)^2 
     \left[  Y_{nm}^{\text{TM/TE}} \csc(k_{z,nm}h)\right] }\;.
\end{multline}
The comparison of the dispersion equations for the mirror- and glide-symmetric structures given in \eqref{eq:disp-mirror} and \eqref{eq:disp-glide1}, respectively, clearly shows that the only difference comes from the factor $(-1)^{n+m}$ in the denominator of~\eqref{eq:disp-glide1}.

The fact that only the sub-unit cell of the the glide-symmetric structure has to be considered is in full correspondence with the sub-unit cell concept introduced in the statement of the \textit{generalized} Floquet's theorem reported in~\cite{highersymmetries2} for periodic structures with higher symmetries. 

The left plot in~\Fig{fig:disp}(d) shows that the results given by the closed-form expression in~\eqref{eq:disp-glide1} are again in good agreement with the data computed by \textit{CST}, which have been obtained by considering the actual unit cell of period $p=2h$. Note that the considered spatial profiles for the glide-symmetric periodic stack, illustrated in~\Fig{fig:disp}(b), are the same as the ones used for the mirror-symmetric periodic stack. The good agreement obtained for both the mirror-symmetric and the glide-symmetric periodic stack of aperture arrays clearly supports the ability of the present closed-form approach to obtain the dispersion diagram of periodic stacked structures even beyond the previous limits of validity of the analytical ECA discussed in~\cite{eca_magazine}. 

It is worth noting here the correlation between the results of transmissivity shown in~Figs.\,\ref{fig:10lay}(b) and~\ref{fig:10lay_glide}(b) and the dispersion behavior of the first Floquet mode in~Figs.\,\ref{fig:disp}(c) and~(d), respectively. The increase of the passband bandwidth observed in~\Fig{fig:10lay_glide}(b) for the glide-symmetric structure completely agrees with the wider bandpass of the first mode in~\Fig{fig:disp}(d) with respect to the one in~\Fig{fig:disp}(c). This effect has been widely reported as one of the advantages of glide-symmetric structures~\cite{JMW2021}, and is shown here to be also a profitable characteristic of stacked FSS's. 

A well-known relevant constraint of most electromagnetic commercial simulators comes from  their inability to provide the attenuation constants of the modes~\cite{APM-2020}. Fortunately, the present approach does not have this limitation since it directly computes the \textit{complex} propagation constants of the Floquet modes. The corresponding values for the normalized attenuation constant ($\alpha_z/k_0$) for the previously analyzed mirror- and glide-symmetric structures  are shown in the right plots of~Figs.\,\ref{fig:disp}(c) and~(d), respectively. For the sake of clarity, only the attenuation constant of the first Floquet mode is shown.  It can be appreciated that,  in the common stopband regions (i.e., from 12.5 to~18.5\,GHz), the attenuation constant of the mirror-symmetric configuration is greater than the one of the glide-symmetric case. This fact is in agreement with the results reported in \cite{glide_dispersion2} for waveguides loaded with holey structures. 

\subsection{Bloch-Floquet Impedance}

\begin{figure}[t]
	\centering
	\subfigure[]{\includegraphics[width= 0.52\columnwidth]{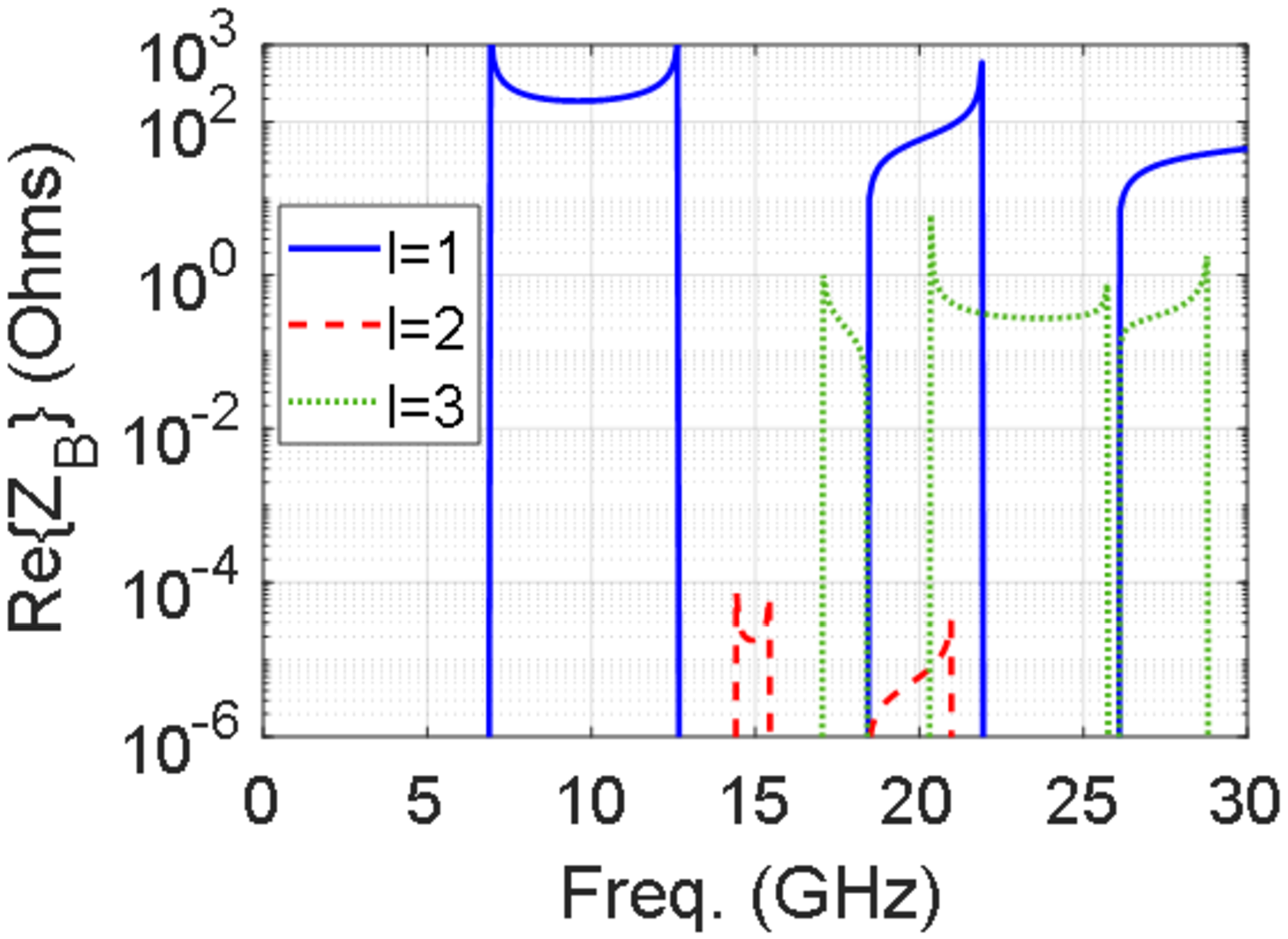}
	}%
	\subfigure[]{\includegraphics[width= 0.5\columnwidth]{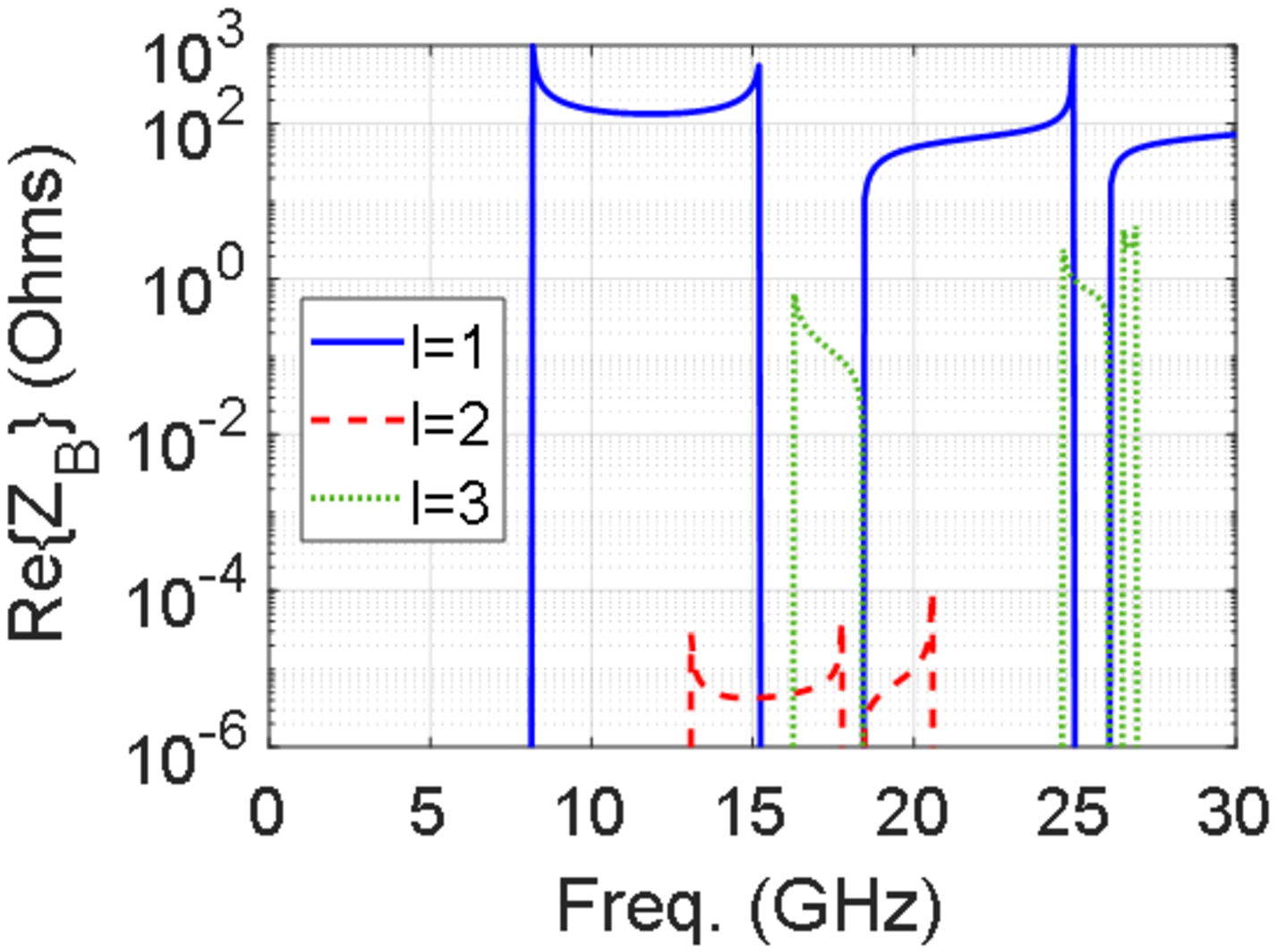}
	} 
	\caption{Real part of the Bloch impedance for various Bloch modes in the  (a)  mirror-symmetric, and (b) glide-symmetric infinite periodic 3-D stacks formed by annular apertures.  Geometrical parameters of the unit cell: $a=3.8$\,mm, $b=4.8$\,mm, $p_x=p_y=p=10$\,mm,  $h=1.575$\,mm, and $\varepsilon_r=2.65$. }
	\label{fig:bloch_impedance}
\end{figure}

As is well known, the introduction of the Bloch-Floquet impedance, given by 
\begin{equation} \label{eq:Bloch_impedance}
Z_\mathrm{B}^\pm = \frac{-2B_p}{A_p-D_p \mp \sqrt{(A_p + D_p)^2 - 4}}
\end{equation}
is very helpful for the study of truncated periodic structures~\cite{pozar}. Unfortunately, most of commercial eigenmode solvers are not able to directly compute the Bloch impedance. The present formulation can overcome this weaknesses and provides accurate information on the Bloch impedance. \Figs{fig:bloch_impedance}(a) and~(b) illustrate
  the real part of the Bloch impedance for the mirror- and glide-symmetric infinite periodic stacks already analyzed in~\Fig{fig:disp}. In the mirror-symmetric structure, after noting that $A_p=D_p$, \eqref{eq:Bloch_impedance} reduces to 
\begin{equation} \label{eq:Bloch_impedance2}
Z_\mathrm{B}^\pm = \pm \dfrac{B_p}{\sqrt{A_p^2 - 1}}\;.
\end{equation}
For the glide-symmetric structure, it is found that the Bloch impedance can alternatively be computed as
\begin{equation} \label{eq:Bloch_impedance_glide}
Z_\mathrm{B}^\pm = \pm\sqrt{\dfrac{A_{p/2}B_{p/2}}{C_{p/2}D_{p/2}}} 
= \pm\sqrt{\dfrac{B_{p/2}}{C_{p/2}}}
\end{equation}
which means that only the sub-unit cell of size $p_z/2$ should be considered in this case.

It can be appreciated that the real part of the Bloch impedance in the first passband region of the $l=1$ mode in~\Fig{fig:bloch_impedance}(a) for the mirror-symmetric configuration (around 195\,$\Omega$ from\,7 to\,12.5\,GHz)  is closer to the free-space impedance ($\eta_0 \approx 377\,\Omega$) than the one in~\Fig{fig:bloch_impedance}(b) for the glide-symmetric structure (around 145\,$\Omega$ from\,8 to\,15\,GHz). This fact explains that the ripples in the passband of the mirror-symmetric structure in~\Fig{fig:10lay} are lower than those for the glide-symmetric structure in~\Fig{fig:10lay_glide}. In addition, the Bloch impedance is progressively smaller in both configurations for the second and third passbands of the first mode ($l=1$), which leads to an increased ripple level for the high frequency passbands, in agreement with the results reported in~\cite{molero_symmetrical1D} for \mbox{1-D} grating stacks. High-order modes ($l=2,3,\ldots$) present a much smaller real part of the Bloch impedance compared to the fundamental mode. Therefore, the resulting mismatching causes these modes to be strongly reflected and hardly transmitted in finite stacks. 

\begin{figure}[t]
	\centering
	\subfigure[]{\includegraphics[width= 0.52\columnwidth]{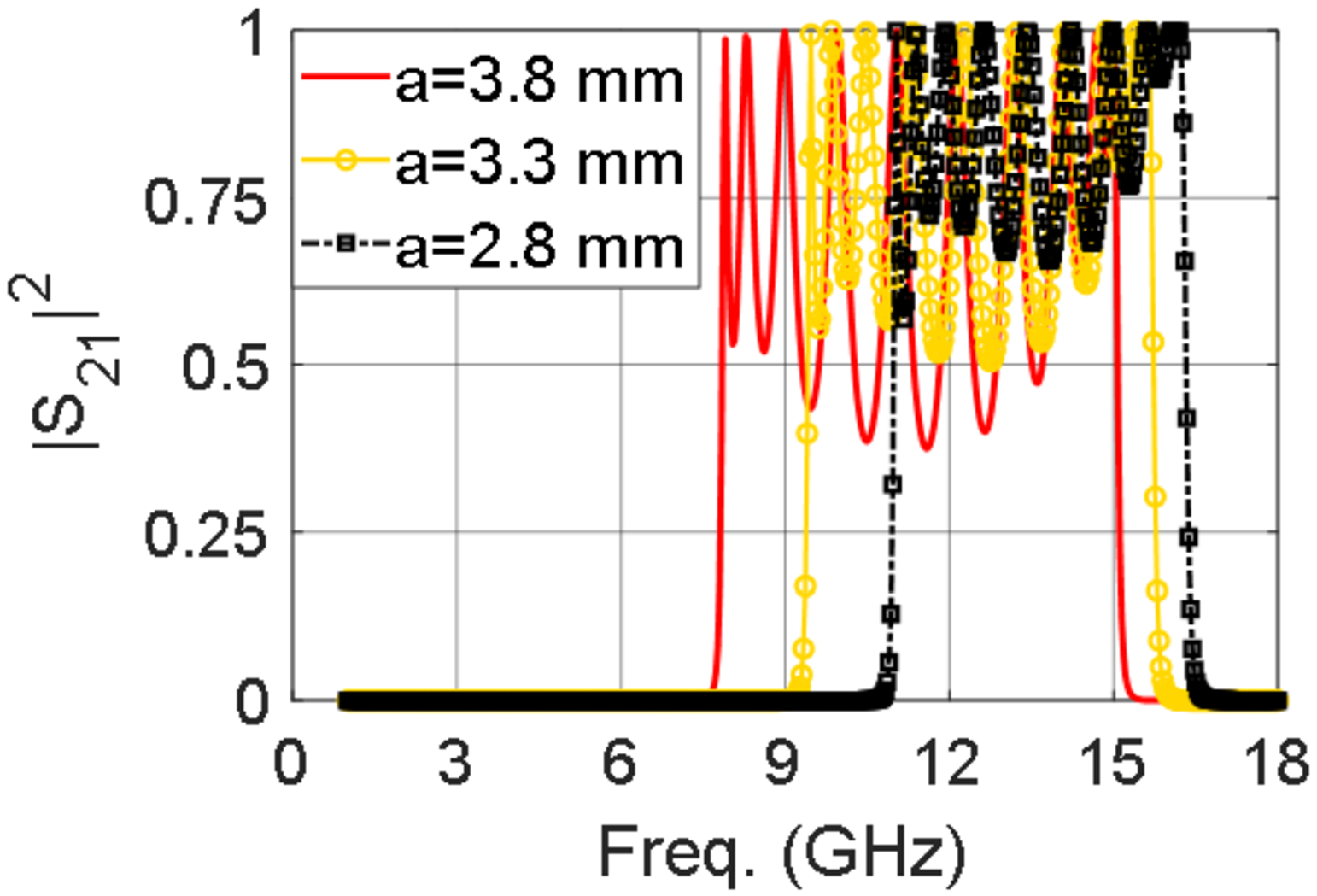}
	} 
	\hspace*{-0.5cm}
	\subfigure[]{\includegraphics[width= 0.48\columnwidth]{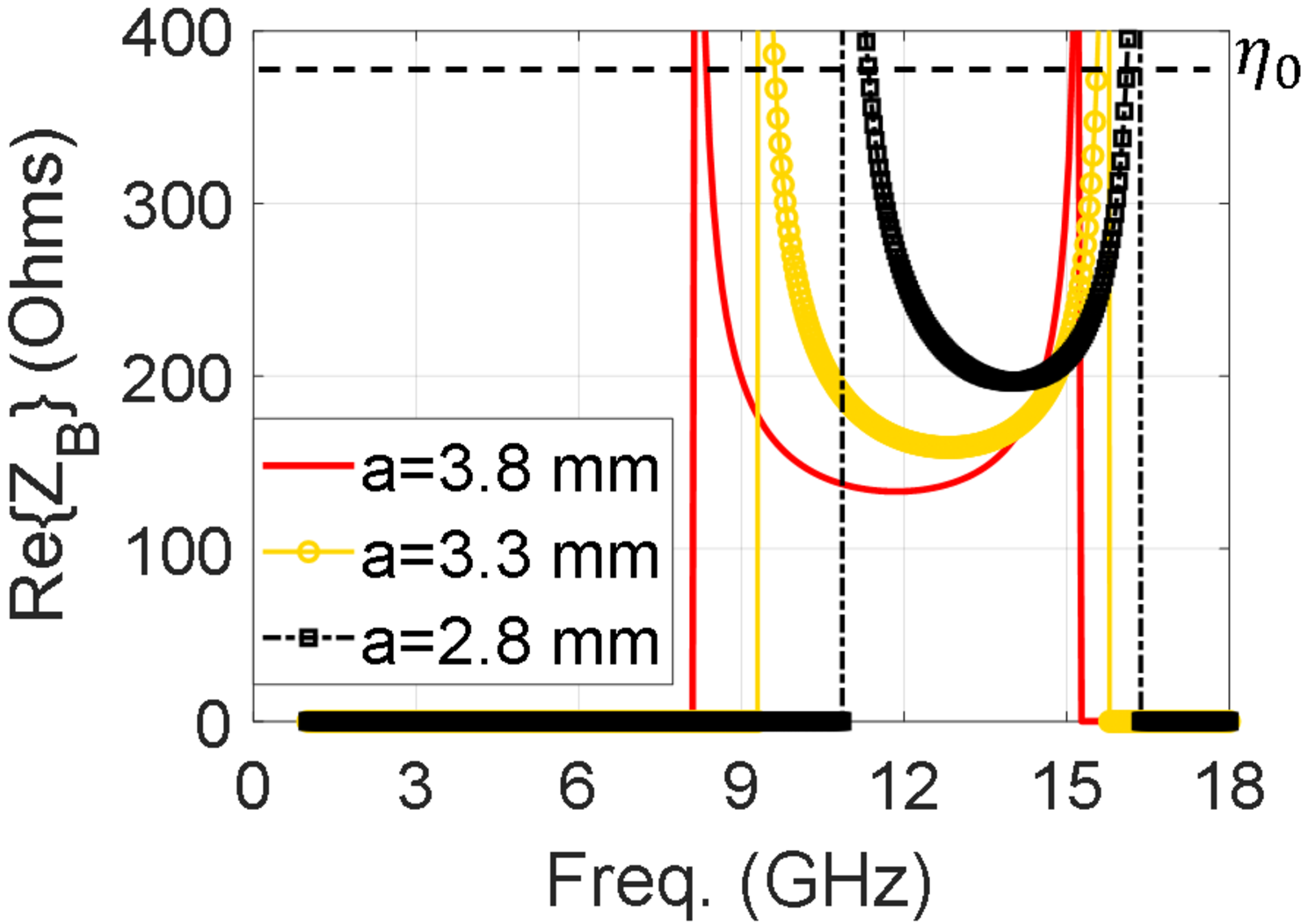}
	} 
	\caption{Reduction of the passband ripple in the glide-symmetric 10-layer  stack formed by annular apertures. (a)~Transmissivity, and (b)~Bloch impedance for different inner radii values, $a$. Geometrical parameters of the unit cell:  $b=4.8$\,mm, $p_x=p_y=p=10$\,mm,  $h=1.575$\,mm, and $\varepsilon_r=2.65$. }
	\label{fig:optimization}
\end{figure}
Bloch impedance can also be used to improve the performance of multilayered FSS structures; in particular, filters and matching layers can be efficiently designed with the proposed circuit approach by adjusting the Bloch impedance of the passband regions to match the free space impedance. As an example, we will show that the passband ripple level of~\Fig{fig:10lay_glide} can be reduced with this simple procedure. Thus, \Fig{fig:optimization}(a) illustrates the variation of the passband ripple level of the glide-symmetric 10-layer stack with glide symmetry as the inner radius of the annular-ring apertures varies. It can be observed that the ripples reduce as the inner radius is smaller; namely, as the width of the annular ring is wider. This effect comes associated with a corresponding increase of the Bloch impedance shown in~\Fig{fig:optimization}(b), which is progressively approaching the value of the free space impedance. As a result, the mismatching between the impedance of the Bloch mode and the free space is reduced and the  transmission is enhanced, although, in this case, at the cost of reducing the passband.

\section{\label{sec:Conc} Conclusion}

This paper presents a rigorous formulation based on the  multi-modal equivalent circuit approach for the analysis and design of stacked structures formed by \mbox{2-D} periodic arrays of arbitrary apertures. It is shown that a key potential of the approach comes from the fact that linear transformations between adjacent layers (rotated, translated and scaled apertures) can be modeled from a purely analytical perspective. This opens new possibilities for the efficient design of polarizers, filters, absorbers, thin matching layers, and other high-frequency devices oriented to wireless communications. As an example, we show the design of a broadband transparent structure formed by annular apertures, a polarization converter, and an absorber formed by rectangular apertures. Good agreement is observed between the present approach and the reference results from \textit{CST} for all the cases under study. Additionally, the present formulation allows for the analysis of glide-symmetric configurations from a circuit perspective. This is a remarkable feature, since the performance of glide-symmetric structures can rarely be described with circuit models due to the strong interaction between adjacent layers. 

Purely analytical results are obtained from the circuit approach as long as the spatial profile of the apertures can be expressed in closed form, regardless of the geometry of the apertures and the applied linear transformations. However, a hybrid approach that combines the use of commercial software and the circuit model can be applied in those cases where the spatial profile cannot be expressed in closed form. This hybrid approach integrates the ability of commercial simulators to deal with arbitrary geometries with the reduced computational effort inherent to the equivalent circuit approach. The hybrid approach is validated with a three-layer stacked structure formed by Jerusalem-cross and bowtie-shaped apertures.

Finally, it is shown that the dispersion properties of infinite periodic stacks can be derived with the proposed formulation. This is a remarkable feature, since most commercial eigenmode software are unable to compute the attenuation constant and Bloch impedance of the modes. At the light of the present results, it is observed that the use of a single spatial profile suffices to compute the dispersion behavior of high-order Floquet modes, as long as the considered profile  resembles the different resonant modes of the aperture. The good agreement shown supports  the  ability  of our  closed-form  approach  to obtain  the  dispersion  diagram  of  periodic  stacked  structures even  beyond  the  previous limits marked  by reference works. 

\appendix

\section{Spatial Aperture Distributions} \label{sec:appendixa}

This appendix gives the analytical expressions of the spatial profiles $\mathbf{E}_a(x,y)$ [see Eq.\eqref{eq:Et}] considered for the computation of the stacked structures. It should be remarked that, although the mathematical form of $\mathbf{E}_a(x,y)$ is assumed to be independent of the dielectric layers inserted in the stacked structure, the information of the dielectric environment is fully accounted for in the ECA by means of the characteristic admittances \eqref{eq:YTM},\eqref{eq:YTE} and wavenumbers \eqref{kzmn} of the transmission lines associated with the Floquet harmonics.

\subsection*{Rectangular Apertures}
Assuming that the tangential electric field at the aperture is oriented as follows: 
\begin{itemize}
\item TM polarization: $\phi=90^\mathrm{o}$, $\mathbf{E}_a(x,y)=E_a(x,y)\,\hat{\mathbf{y}}$
\item TE polarization: \;$\phi=0^\mathrm{o}$,\;\; $\mathbf{E}_a(x,y)=E_a(x,y)\,\hat{\mathbf{x}}$\,,
\end{itemize}
 the following spatial profiles can be considered in the case of a rectangular aperture of dimensions $a \times b$ \cite{basis_functions}: 
\begin{align} \label{spatialprofile_cos}
{E}_{a1}(x,y)&  \propto \bigg[\cos\!\Big(\frac{\pi x}{a} \Big)\, \mathrm{rect}\Big( \frac{x}{a} \Big)\bigg] \, \mathrm{rect}\Big( \frac{y}{b} \Big) \\
\label{spatialprofile_cos_sqrt}
	{E}_{a2}(x,y) & \propto \bigg[\frac{\cos (\frac{\pi x}{a})}{\sqrt{1-(2x/a)^2}} \, \mathrm{rect}\Big( \frac{x}{a} \Big) \bigg]\, \mathrm{rect}\Big( \frac{y}{b} \Big)
\end{align}
where  the rectangle function is defined as
\begin{equation}
	\mathrm{rect}\Big(\frac{x}{u}\Big)= 
	\begin{cases}
	1\,, & -u/2 \leq x \leq u/2 \\
	0\,, & \text{otherwise}\;.
	\end{cases}
\end{equation}
A further study demonstrated that the spatial distribution \eqref{spatialprofile_cos} offer more accurate results when the aperture size is narrow, while \eqref{spatialprofile_cos_sqrt} can be employed for bigger apertures.

The Fourier transforms of these spatial profiles can be expressed in closed form as
\begin{equation}  \label{Ea1_rectangular}
{\tilde{E}}_{a1}(k_{xn},k_{ym}) \propto \Bigg[ \dfrac{\cos\! \Big( k_{xn}\frac{a}{2} \Big)}{\sqrt{(\pi/ a)^2 - k^2_{xn}}}    \Bigg]  \,
\frac{\sin\!\Big( k_{ym}\frac{b}{2}\Big)}{k_{ym}}
\end{equation}
\begin{multline} \label{tildespatialprofile_cos_sqrt}
{\tilde{E}}_{a2}(k_{xn},k_{ym}) \propto \bigg[ J_0\Big(  \Big|k_{xn}+\frac{\pi}{a} \Big|\frac{a}{2} \Big) \\ +
J_0\Big( \Big|k_{xn} - \frac{\pi}{a} \Big|\frac{a}{2} \Big) \bigg] \,
\frac{\sin\!\Big(k_{ym}\frac{b}{2} \Big)}{k_{ym}}
\end{multline}
with
$J_0(\cdot)$ being the Bessel function of the first kind and order zero.  The transformer turn ratios $N_{nm}^\mathrm{TM/TE}$ can be directly calculated with~\eqref{Ea1_rectangular} and~\eqref{tildespatialprofile_cos_sqrt} from [\citenum{evenodd2}, Eqs.\,(6) and (7)].

\subsection*{Annular Apertures}
The spatial profile of annular apertures shares a great similarity with of rectangular apertures. In the case of annular apertures, field variations occur in the azimuth direction $\phi$, rather than in $x$ or $y$ directions. According to \cite{fss_eca2006, beruete_anular}, the following aperture field can be considered in an annular ring whose inner and outer radii are $a$ and $b$, respectively,
\begin{equation}\label{eq:Ea_anular}
\mathbf{E}_a(\phi) \propto \cos\big[ l (\phi - \phi_0)\big] \,  \hat{\boldsymbol{\rho}}
\end{equation}
where $\phi_0$ is the reference azimuth angle, and $l$ stands for the the order of the considered mode. As the variation in the radial direction has been suppressed, this approximation is valid as long as the slot width is narrow ($1 \leq b/a \lessapprox 1.5$). For the fundamental mode ($l=1$),  the expressions
for the transformer turn ratios $N_{nm}^\mathrm{TM/TE}$ are found in [\citenum{beruete_anular}, Eqs.\,(5) and (6)].

\ifCLASSOPTIONcaptionsoff
  \newpage
\fi



%

\end{document}